\documentclass[aps,prb,reprint,superscriptaddress]{revtex4-2}
\usepackage{amsfonts}
\usepackage{amssymb}
\usepackage{amsmath}
\usepackage{graphicx}
\usepackage{epsfig}
\usepackage{array}
\usepackage{braket}
\usepackage{multirow}
\usepackage[table,xcdraw]{xcolor}
\usepackage[colorlinks=true, citecolor=blue, urlcolor=blue, linkcolor=red]{hyperref}

\begin{document}

\title{Topological classification and edge states of magnons in honeycomb
ferromagnets}
\author{Youwen Wang}
\altaffiliation{These authors contributed equally to this work.}
\author{Qiutong Wang}
\altaffiliation{These authors contributed equally to this work.}
\author{Qingjun Tong}
\author{Ci Li}
\email{lici@hnu.edu.cn}
\affiliation{School of Physics and Electronics, Hunan University, Changsha
410082, China}

\begin{abstract}
We study the topological classification and related edge states of magnons
in ferromagnets on honeycomb that can be described by a class of
single-particle bosonic Bogoliubov-de Gennes (BdG) models. Both single layer
and bilayer situations are considered. The calculations show that the
existence and related topologies of these edge states are well captured by a
class of non-Hermitian single or coupled Su-Schrieffer-Heeger chains models $%
H(k_y)$ parameterized by momentum $k_y$, where the edge states can appear as
the ground state for some cases. Interestingly, although the eigenproblem of
bosonic BdG models is equivalent to the one of non-Hermitian systems, the
conventional bulk-edge correspondence for Hermitian systems is partially
valid. The influence of Dzyaloshinskii-Moriya interactions between next
nearest-neighbor spins are also discussed, which break the time-reversal
symmetry and lead to a straight connection between edge states for magnonic
systems and non-zero Chern number of non-Hermitian bulk two-dimensional
systems.
\end{abstract}

\maketitle

%\altaffiliation{These authors contributed equally to this work.}

%\altaffiliation{These authors contributed equally to this work.}

%\affiliation{School of Physics and Electronics, Hunan University, Changsha
%410082, China}

%\author{Wang Yao}
%\email{wangyao@hku.hk}

%\affiliation{Department of Physics, The University of Hong Kong, Hong Kong,
%China} 
%\affiliation{HKU-UCAS Joint Institute of Theoretical and Computational Physics at
%Hong Kong, China}

\section{Introduction}

As one of the most significant quasiparticles, the magnon-representing the
collective excitation of the spin structure of electrons in a crystal
lattice \cite{Blo}-has garnered considerable attention from researchers in
condensed matter physics for its potential applications in information
encoding and processing \cite{Kai,Mun,Wee,Hil,Xia,Yan,You}, also spintronics 
\cite{Yan,Ser,Wun,Mcc,Lou,Moo,Moo1,Sal,Kis}. In recent years, substantial
advancements in topological materials have provided new insights into the
geometry and topology of magnons \cite{Mcc}. Berry phases associated with
magnetic dynamics can lead to observable effects in heat and spin transport 
\cite{Lee,Tok,Mur,Mur1,Ong,Ong1,DX,Zyu,Tse,Ran,Mer}, while analogs of
topological insulators and semimetals can emerge within magnon band
structures due to intrinsic magnetic couplings \cite%
{Lou,Moo,Kis,Ran,Mer,Fie,Col,Dag,Hir,Los,Los1,Wu}. These characteristics
position magnons as crucial elements in exploring the interplay between
magnetic symmetries and topology \cite{Kat}, influencing topological
transitions through magnetic fields, and investigating the impact of
interactions on topological bands. Moreover, magnons hold the promise of
generating topologically protected spin currents at interfaces, underscoring
their vital role in advancing the field.

In the equivalent wave picture of quantum mechanics, a magnon can be viewed
as a quantized spin wave with a fixed amount of energy and lattice momentum 
\cite{Blo}, indicating that magnons exhibit bosonic behavior. It has been
demonstrated that magnonic systems arising from lattice models of electronic
spin systems can be effectively described by single-particle bosonic
Bogoliubov-de Gennes (BdG) Hamiltonians \cite{Mcc,Kat,Aue,Nolt,Sim,Kaz},
whose eigenproblems equal to those of related non-Hermitian systems \cite%
{Mcc,Kat,Sat}. This connection allows for the topological classification of
magnon systems to be seamlessly linked to the well-established non-Hermitian
topological classification theory \cite{Kat,Sat}.

\begin{figure}[tbp]
\begin{center}
\includegraphics[width=0.48\textwidth]{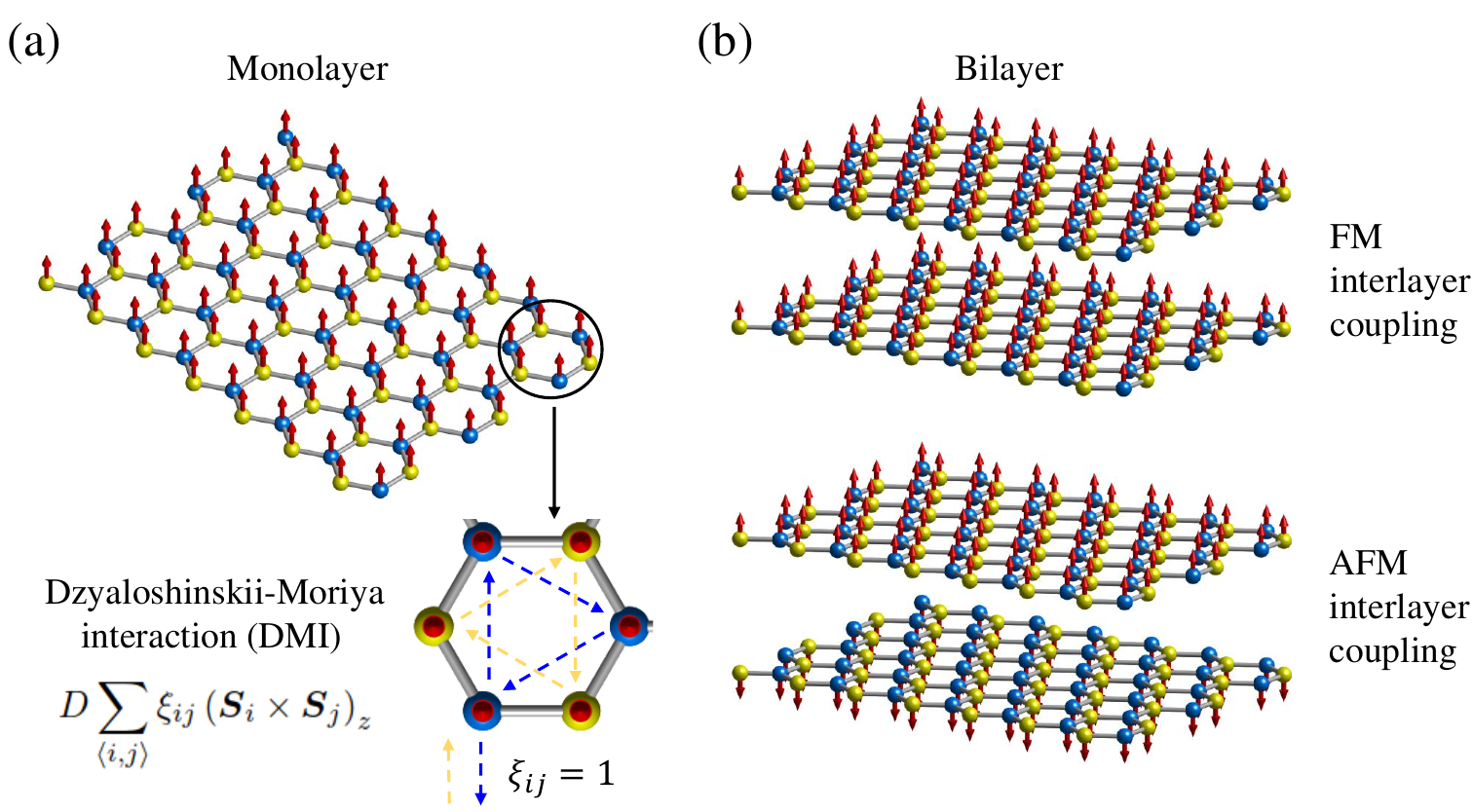}
\end{center}
\caption{(Color online) (a) Schematics of monolayer ferromagnets and
possible form of Dzyaloshinskii-Moriya interaction in honeycomb lattice
structure. The localized spins are represented by red arrows and point along
the z axis. (b) Similar plots for bilayer ferromagnets with different
interlayer couplings.}
\label{fig1}
\end{figure}

However, compared to previous studies that primarily focus on topological
bulk magnon systems in honeycomb antiferromagnets \cite%
{Kis,DX,Zyu,Mer,Hir,Los} and monolayer honeycomb ferromagnets \cite%
{Tse,Los1,Wu,Kat,Bal,Tse1}, the detailed topological classification and
related edge states of magnon systems in hexagonal ferromagnets, especially
bilayer ferromagnets or monolayer ferromagnets with different boundary
conditions, have not been thoroughly explored. In this paper, we
systematically investigate the topological classification and related edge
states of magnon systems originating from two-dimensional (2D) ferromagnets
with localized spin moments arranged on a honeycomb lattice in the~$xy$%
-plane, as illustrated in Fig. \ref{fig1}. Both monolayer and bilayer
configurations are considered.

For the monolayer case, since there are no double annihilation or creation
operators in the related BdG models, the conventional bulk-edge
correspondence for Hermitian systems remains effective. Non-trivial edge
states can emerge even without the Dzyaloshinskii-Moriya interaction (DMI) 
\cite{D,M} under certain open boundary conditions. The existence and
topologies of these edge states are well captured by a class of
non-Hermitian coupled Su-Schrieffer-Heeger (SSH) chain models~$H(k_{y})$%
~parameterized by momentum~$k_{y}${}, where edge states can appear as the
ground state for some boundary conditions. When DMI is applied, the bulk BdG
Hamiltonian typically becomes topologically non-trivial with a non-zero
Chern number. In this scenario, it is well-known that the related edge
states can be connected to the Chern number through their winding numbers 
\cite{Hat,Hat1,Wan}. For the bilayer case, the stacking form (AA or AB) and
the type of interlayer coupling, i.e., ferromagnetic (FM) or
antiferromagnetic (AFM), can significantly alter the Chern number for magnon
bands. Additionally, the different boundary conditions for different layers
play a crucial role in the distribution of edge states in momentum space.

The paper is arranged in the following: In section \ref{2}, we give a brief
description of ferromagnets and antiferromagnets on honeycomb lattices based
on the Heisenberg Hamiltonian. In section \ref{3}, we show the topological
classification and related edge states for magnons in monolayer ferromagnets
on honeycomb lattices, both BdG Hamiltonians with periodic and open boundary
conditions are discussed in detail. In section \ref{4}, we concentrate on
the situation for bilayer ferromagnets on honeycomb lattices. Finally, we
give a summary.

\section{General description for magnons in ferromagnets and
antiferromagnets on honeycomb lattices}

\label{2}

\subsection{Ferromagnetic (FM) Heisenberg Hamiltonian}

At first, we take a simple Heisenberg Hamiltonian of a monolayer ferromagnet
on honeycomb lattice \cite{Mcc,Aue,Nolt}%
\begin{equation}
H=-J\sum_{\left\langle i,j\right\rangle }\boldsymbol{S}_{i}\cdot \boldsymbol{%
S}_{j}+D\sum_{\left\langle \left\langle i,j\right\rangle \right\rangle }\xi
_{ij}\left( \boldsymbol{S}_{i}\times \boldsymbol{S}_{j}\right) _{z},
\label{FM}
\end{equation}%
as an example. $J>0$ is the nearest neighbor (NN) spin exchange. The latter
term with real number $D$ means DMI between next nearest-neighbor (NNN)
spins since the NN DMI vanishes in honeycomb lattices \cite{DX,D,M}. In
principle, the sign convention $\xi _{ij}$ can be chosen artificially. In
the rest of paper, we use the convention shown in Fig. \ref{fig1}, which
remains zero net flux in one unit cell and well preserves the space
translational symmetry \cite{Che}. Here $i,j$ denotes the unit cell of
honeycomb lattice. If we assume there is a N\'{e}el order in the direction
perpendicular to the lattice plane, i.e., $+z$ direction, the
Holstein-Primakoff (HP) transformation \cite{Mcc,Aue,Nolt} that change spins
to bosonic operators (magnons) becomes%
\begin{eqnarray}
S_{i}^{+} &=&S_{i}^{x}+iS_{i}^{y}=\sqrt{2S-f_{i}^{\dagger }f_{i}}f_{i}, \\
S_{i}^{-} &=&S_{i}^{x}-iS_{i}^{y}=f_{i}^{\dagger }\sqrt{2S-f_{i}^{\dagger
}f_{i}},  \notag \\
S_{i}^{z} &=&S-f_{i}^{\dagger }f_{i},  \notag
\end{eqnarray}%
where $f_{i}\left( f_{i}^{\dagger }\right) $ represents the bosonic
annihilation (production) operator and $S_{j}^{l}\left( l=+,-,z\right) $ has
the same form as $S_{i}^{l}$ since they have the same spin direction ($+z$).
Considering the large $S$ limit, the Taylor expansion gives%
\begin{equation}
S_{i}^{+}\approx \left( \sqrt{2S}-\frac{f_{i}^{\dagger }f_{i}}{2\sqrt{2S}}%
\right) f_{i},S_{i}^{-}\approx f_{i}^{\dagger }\left( \sqrt{2S}-\frac{%
f_{i}^{\dagger }f_{i}}{2\sqrt{2S}}\right) .
\end{equation}%
Only keeping the linear order, the straightforward calculation shows%
\begin{eqnarray*}
\boldsymbol{S}_{i}\cdot \boldsymbol{S}_{j} &=&\frac{1}{2}\left(
S_{i}^{+}S_{j}^{-}+S_{i}^{-}S_{j}^{+}\right) +S_{i}^{z}S_{j}^{z} \\
&\approx &S\left( f_{i}f_{j}^{\dagger }+f_{i}^{\dagger }f_{j}\right)
-S\left( f_{i}^{\dagger }f_{i}+f_{j}^{\dagger }f_{j}\right) +S^{2},
\end{eqnarray*}%
\begin{eqnarray}
\left( \boldsymbol{S}_{i}\times \boldsymbol{S}_{j}\right) _{z}
&=&S_{i}^{x}S_{j}^{y}-S_{i}^{y}S_{j}^{x}=\frac{i}{2}\left(
S_{i}^{+}S_{j}^{-}-S_{i}^{-}S_{j}^{+}\right)   \notag \\
&\approx &iS\left( f_{i}f_{j}^{\dagger }-f_{i}^{\dagger }f_{j}\right) .
\end{eqnarray}%
Since the two spins $\boldsymbol{S}_{i}$ and $\boldsymbol{S}_{j}$ with an
arbitrary angle between them can be expressed as \cite{Zyu}%
\begin{equation}
\boldsymbol{S}_{i}=\left( 
\begin{array}{c}
\frac{1}{2}\left( S_{i}^{+}+S_{i}^{-}\right)  \\ 
\frac{1}{2i}\left( S_{i}^{+}-S_{i}^{-}\right)  \\ 
S_{i}^{z}%
\end{array}%
\right) ,\boldsymbol{S}_{j}=R_{\theta }R_{\phi }\left( 
\begin{array}{c}
\frac{1}{2}\left( S_{j}^{+}+S_{j}^{-}\right)  \\ 
\frac{1}{2i}\left( S_{j}^{+}-S_{j}^{-}\right)  \\ 
S_{j}^{z}%
\end{array}%
\right) ,
\end{equation}%
with rotation matrices%
\begin{equation}
R_{\theta }=\left( 
\begin{array}{ccc}
\cos \theta  & 0 & -\sin \theta  \\ 
0 & 1 & 0 \\ 
\sin \theta  & 0 & \cos \theta 
\end{array}%
\right) ,R_{\phi }=\left( 
\begin{array}{ccc}
\cos \phi  & \sin \phi  & 0 \\ 
-\sin \phi  & \cos \phi  & 0 \\ 
0 & 0 & 1%
\end{array}%
\right) .
\end{equation}%
The difference between $+z$ and $-z$ requires $\theta =\pi $ and $\phi =0$,
so the HP transformation for the spin $\boldsymbol{S}_{j}$ with $-z$
direction requires 
\begin{equation}
\frac{1}{2}\left( S_{j}^{+}+S_{j}^{-}\right) =-S_{j}^{x},\frac{1}{2i}\left(
S_{j}^{+}-S_{j}^{-}\right) =S_{j}^{y},S_{j}^{z}=-S+f_{j}^{\dagger }f_{j}.
\end{equation}%
and the nearest-neighbor spins with the same direction ($-z$) gives%
\begin{eqnarray*}
\boldsymbol{S}_{i}\cdot \boldsymbol{S}_{j} &=&\frac{1}{2}\left(
S_{i}^{+}S_{j}^{-}+S_{i}^{-}S_{j}^{+}\right) +S_{i}^{z}S_{j}^{z} \\
&\approx &S\left( f_{i}f_{j}^{\dagger }+f_{i}^{\dagger }f_{j}\right)
-S\left( f_{i}^{\dagger }f_{i}+f_{j}^{\dagger }f_{j}\right) +S^{2},
\end{eqnarray*}%
\begin{eqnarray}
\left( \boldsymbol{S}_{i}\times \boldsymbol{S}_{j}\right) _{z}
&=&S_{i}^{x}S_{j}^{y}-S_{i}^{y}S_{j}^{x}=-\frac{i}{2}\left(
S_{i}^{+}S_{j}^{-}-S_{i}^{-}S_{j}^{+}\right)   \notag \\
&\approx &-iS\left( f_{i}f_{j}^{\dagger }-f_{i}^{\dagger }f_{j}\right) .
\end{eqnarray}

\subsection{Antiferromagnetic (AFM) Heisenberg model}

Following the same assumption before, a simple Heisenberg Hamiltonian of a
monolayer antiferromagnet on honeycomb lattice can be expressed as \cite%
{Mcc,Aue,Nolt}

\begin{equation}
H=J\sum_{\left\langle i,j\right\rangle }\boldsymbol{S}_{i}\cdot \boldsymbol{S%
}_{j}+D\sum_{\left\langle \left\langle i,j\right\rangle \right\rangle }\xi
_{ij}\left( \boldsymbol{S}_{i}\times \boldsymbol{S}_{j}\right) _{z},
\label{AFM}
\end{equation}%
Now the N\'{e}el order requires that the nearest-neighbor spins should have
the opposite directions ($+z$ and $-z$). Still keeping the linear order and
dropping all higher order terms after the HP transformation, we have%
\begin{eqnarray}
\boldsymbol{S}_{i}\cdot \boldsymbol{S}_{j} &=&-\frac{1}{2}\left(
S_{i}^{+}S_{j}^{+}+S_{i}^{-}S_{j}^{-}\right) +S_{i}^{z}S_{j}^{z} \\
&\approx &-S\left( f_{i}f_{j}+f_{i}^{\dagger }f_{j}^{\dagger }\right)
+S\left( f_{i}^{\dagger }f_{i}+f_{j}^{\dagger }f_{j}\right) -S^{2},  \notag
\end{eqnarray}%
\begin{eqnarray*}
\left( \boldsymbol{S}_{i}\times \boldsymbol{S}_{j}\right) _{z}
&=&S_{i}^{x}S_{j}^{y}-S_{i}^{y}S_{j}^{x}=\frac{i}{2}\left(
S_{i}^{-}S_{j}^{-}-S_{i}^{+}S_{j}^{+}\right) \\
&\approx &-iS\left( f_{i}f_{j}-f_{i}^{\dagger }f_{j}^{\dagger }\right) .
\end{eqnarray*}

\subsection{Other terms may exist in the Heisenberg Hamiltonian}

For a general FM or AFM system, the description based on the Heisenberg
model can be very complex by including other terms such as NNN spin exchange
or anisotropic Zeeman term which breaks the translational symmetry. In this
paper, we only consider two simple extra terms that may detune the energy of
the system. One is the easy-axis anisotropic term that usually used in
collinear antiferromagnets to keep the N\'{e}el vector in the $z$ direction 
\cite{DX}, which can be expressed as%
\begin{equation}
\mathcal{K}\sum_{i}S_{i,z}^{2}\approx \mathcal{K}S\sum_{i}\left(
S-2f_{i}^{\dagger }f_{i}\right) ,
\end{equation}%
for both spin up $\left( +z\right) $ and down $\left( -z\right) $, with $%
\mathcal{K}<0$ in the linear order after the HP transformation. It normally
does not change the symmetry of the system. The other is the isotropic
Zeeman term that generally has the form%
\begin{eqnarray}
&&\frac{g\mu _{B}}{\hbar }\boldsymbol{B}\cdot \sum_{i}\boldsymbol{S}_{i} \\
&\approx &\left\{ 
\begin{array}{c}
\frac{g\mu _{B}}{\hbar }B_{z}\sum_{i}\left( S-f_{i}^{\dagger }f_{i}\right) ,%
\text{spin up }\left( +z\right) \\ 
-\frac{g\mu _{B}}{\hbar }B_{z}\sum_{i}\left( S-f_{i}^{\dagger }f_{i}\right) ,%
\text{spin down }\left( -z\right)%
\end{array}%
\right. ,  \notag
\end{eqnarray}%
in the linear order after the HP transformation, with $g$ is the $g$-factor
and $\mu _{B}$ is the Bohr magneton \cite{Tok,Mer}, where the isotropy comes
from the isotropic magnetic field $B_{z}$ that perpendicular to the $xy$
plane. It is normally used to spilt the energy of different spins.

\subsection{Diagonalization and topological classification of
non-interacting bosonic quadratic Hamiltonians}

If we only consider the linear order after the HP transformation. For both
FM and AFM cases or other more complicated cases, we can always rewrite the
Heisenberg Hamiltonian into the standard non-interacting bosonic quadratic
form \cite{Mcc,Kat,Sim,Kaz,Sat}, which is%
\begin{equation}
H=\frac{1}{2}\left( 
\begin{array}{cc}
\boldsymbol{\beta }^{\dagger } & \boldsymbol{\beta }%
\end{array}%
\right) \left( 
\begin{array}{cc}
h & \Delta \\ 
\Delta ^{\ast } & h^{T}%
\end{array}%
\right) \left( 
\begin{array}{c}
\boldsymbol{\beta } \\ 
\boldsymbol{\beta }^{\dagger }%
\end{array}%
\right) ,
\end{equation}%
with $\boldsymbol{\beta }=\left( \beta _{1},\beta _{2},...,\beta _{N}\right) 
$ and $\Delta ^{T}=\Delta $ since Bose statistics ($\beta _{m}\beta
_{n}=\beta _{n}\beta _{m}$). The $2N\times 2N$ Hermitian matrix can be seen
as the bosonic BdG Hamiltonian \cite{Kat,Sim,Kaz}%
\begin{equation}
H_{BdG}=\left( 
\begin{array}{cc}
h & \Delta \\ 
\Delta ^{\ast } & h^{T}%
\end{array}%
\right) .
\end{equation}%
The bosonic commutation requires%
\begin{equation}
\left( 
\begin{array}{cc}
\boldsymbol{\beta }^{\dagger } & \boldsymbol{\beta }%
\end{array}%
\right) \sum\nolimits_{z}\left( 
\begin{array}{c}
\boldsymbol{\beta } \\ 
\boldsymbol{\beta }^{\dagger }%
\end{array}%
\right) =-N,
\end{equation}%
with $\sum\nolimits_{z}=\sigma _{z}\otimes \mathit{I}_{N}$, $\sigma _{z}$
the $2\times 2$ Pauli matrix and $\mathit{I}_{N}$ the $N\times N$ identity
matrix. If $H$ can be diagonalized, i.e.,%
\begin{eqnarray}
&&\left( 
\begin{array}{cc}
\boldsymbol{\beta }^{\dagger } & \boldsymbol{\beta }%
\end{array}%
\right) \left( Q^{\dagger }\right) ^{-1}Q^{\dagger }H_{BdG}QQ^{-1}\left( 
\begin{array}{c}
\boldsymbol{\beta } \\ 
\boldsymbol{\beta }^{\dagger }%
\end{array}%
\right)  \notag \\
&=&\left( 
\begin{array}{cc}
\boldsymbol{\beta }^{\dagger } & \boldsymbol{\beta }%
\end{array}%
\right) \left( Q^{\dagger }\right) ^{-1}\Lambda Q^{-1}\left( 
\begin{array}{c}
\boldsymbol{\beta } \\ 
\boldsymbol{\beta }^{\dagger }%
\end{array}%
\right) ,
\end{eqnarray}%
where $\Lambda $ is a diagonal matrix. The quasi-bosons should satisfy%
\begin{equation}
\left( 
\begin{array}{cc}
\boldsymbol{\beta }^{\dagger } & \boldsymbol{\beta }%
\end{array}%
\right) \left( Q^{\dagger }\right) ^{-1}\sum\nolimits_{z}Q^{-1}\left( 
\begin{array}{c}
\boldsymbol{\beta } \\ 
\boldsymbol{\beta }^{\dagger }%
\end{array}%
\right) =-N,
\end{equation}%
it leads to $Q^{\dagger }\sum\nolimits_{z}Q=\sum\nolimits_{z}$, that implies
the transformation matrix $Q$ is symplectic (i.e., nonunitary), which cannot
happen in the fermionic systems since%
\begin{equation}
\left( 
\begin{array}{cc}
\boldsymbol{\beta }^{\dagger } & \boldsymbol{\beta }%
\end{array}%
\right) \left( 
\begin{array}{c}
\boldsymbol{\beta } \\ 
\boldsymbol{\beta }^{\dagger }%
\end{array}%
\right) =N,
\end{equation}%
for fermionic cases. Using $Q^{\dagger
}=\sum\nolimits_{z}Q^{-1}\sum\nolimits_{z}$, one can easily find that $%
\sum\nolimits_{z}Q^{-1}\sum\nolimits_{z}H_{BdG}Q=\Lambda $, giving $%
Q^{-1}\sum\nolimits_{z}H_{BdG}Q=\sum\nolimits_{z}\Lambda $. This implies
that the eigenvalues of $H_{\sigma BdG}\equiv \sum\nolimits_{z}H_{BdG}$ will
determine the spectrum of the system and the right eigenvectors of it will
determine the transformation matrix $Q$ \cite{Kat,Sim,Kaz,Sat}.

For such a Hamiltonian $H$ that has the periodic boundary condition (PBC),
we can investigate the physical properties of it in momentum space. Assuming
that the unit cell of the system has $l$ types of modes (atoms), then it
leads to%
\begin{equation}
H\left( \boldsymbol{k}\right) =\frac{1}{2}\left( 
\begin{array}{cc}
\boldsymbol{\beta }_{\boldsymbol{k}}^{\dagger } & \boldsymbol{\beta }_{-%
\boldsymbol{k}}%
\end{array}%
\right) \left( 
\begin{array}{cc}
h\left( \boldsymbol{k}\right) & \Delta \left( \boldsymbol{k}\right) \\ 
\Delta ^{\ast }\left( -\boldsymbol{k}\right) & h^{\ast }\left( -\boldsymbol{k%
}\right)%
\end{array}%
\right) \left( 
\begin{array}{c}
\boldsymbol{\beta }_{\boldsymbol{k}} \\ 
\boldsymbol{\beta }_{-\boldsymbol{k}}^{\dagger }%
\end{array}%
\right) ,
\end{equation}%
with $\boldsymbol{\beta }_{\boldsymbol{k}}=\left( \beta _{\boldsymbol{k}%
}^{1},\beta _{\boldsymbol{k}}^{2},...,\beta _{\boldsymbol{k}}^{l}\right) $.
The analysis above in the real space is still effective here and we can
define%
\begin{equation}
H_{BdG}\left( \boldsymbol{k}\right) =\left( 
\begin{array}{cc}
h\left( \boldsymbol{k}\right) & \Delta \left( \boldsymbol{k}\right) \\ 
\Delta ^{\ast }\left( -\boldsymbol{k}\right) & h^{\ast }\left( -\boldsymbol{k%
}\right)%
\end{array}%
\right) ,  \label{G_BdG}
\end{equation}%
with related $H_{\sigma BdG}\left( \boldsymbol{k}\right) \equiv
\sum\nolimits_{z}^{k}H_{BdG}\left( \boldsymbol{k}\right) $, $%
\sum\nolimits_{z}^{k}=\sigma _{z}\otimes \mathit{I}_{l}$. Following the
topological classification defined in Ref. \cite{Sat}, the symmetry $%
H_{\sigma BdG}\left( \boldsymbol{k}\right) $ respects can be time-reversal
symmetry (TRS)%
\begin{eqnarray}
T_{+}H_{\sigma BdG}^{\ast }\left( \boldsymbol{k}\right) T_{+}^{-1}
&=&H_{\sigma BdG}\left( -\boldsymbol{k}\right) , \\
T_{+}T_{+}^{\ast } &=&\pm 1,T_{+}T_{+}^{\dagger }=1,  \notag
\end{eqnarray}%
and pesudo-Hermiticity (pH)%
\begin{equation}
\eta H_{\sigma BdG}^{\dagger }\left( \boldsymbol{k}\right) \eta
^{-1}=H_{\sigma BdG}\left( \boldsymbol{k}\right) ,\eta ^{2}=1,\eta =\eta
^{\dagger }.
\end{equation}%
Since $H_{\sigma BdG}^{\dagger }\left( \boldsymbol{k}\right) =\left[
\sum\nolimits_{z}^{k}H_{BdG}\left( \boldsymbol{k}\right) \right] ^{\dagger
}=H_{BdG}^{\dagger }\left( \boldsymbol{k}\right)
\sum\nolimits_{z}^{k}=H_{BdG}\left( \boldsymbol{k}\right)
\sum\nolimits_{z}^{k}$, the pH can always be satisfied with $\eta
=\sum\nolimits_{z}^{k}$. If the real gap exists for $H_{\sigma BdG}\left( 
\boldsymbol{k}\right) $, the related Chern number in even spatial dimensions 
$d=2n$ can be defined as \cite{Sat}%
\begin{equation}
C_{n}=\frac{1}{n!}\left( \frac{i}{2\pi }\right) ^{n}\int_{BZ^{d}}tr\mathcal{F%
}^{n},
\end{equation}%
the trace is taken over the occupied bands. $\mathcal{F}$ is the non-Abelian
Berry curvature. For the 2D case, i.e., $n=1$, it returns to the general form%
\begin{equation}
C_{1}=\frac{1}{2\pi }\sum_{p<0}\int_{BZ^{d}}\Omega _{p,p}^{k_{x}k_{y}}%
\mathrm{d}\boldsymbol{k}.
\end{equation}%
$p<0$ means the occupied bands.

Finally, we can conclude the topological classification of arbitrary $%
H_{\sigma BdG}\left( \boldsymbol{k}\right) $ in 2D situation as \cite{Sat}:

I. Class $\mathcal{A}$ (without any symmetry except pH), there maybe a $%
%TCIMACRO{\U{2124} }%
%BeginExpansion
\mathbb{Z}
%EndExpansion
$ topological invariant.

ll. Class $\mathcal{AI}$ with $\eta _{+}$ (both TRS and pH exist with $%
T_{+}T_{+}^{\ast }=1,\left[ T_{+},\eta \right] =0$), which is always
topological trivial.

III. Class $\mathcal{AI}$ with $\eta _{-}$ (both TRS and pH exist with $%
T_{+}T_{+}^{\ast }=1,\left\{ T_{+},\eta \right\} =0$), there maybe a $%
%TCIMACRO{\U{2124} }%
%BeginExpansion
\mathbb{Z}
%EndExpansion
$ topological invariant.

lV. Class $\mathcal{AII}$ with $\eta _{+}$ (both TRS and pH exist with $%
T_{+}T_{+}^{\ast }=-1,\left[ T_{+},\eta \right] =0$), there maybe a $%
%TCIMACRO{\U{2124} }%
%BeginExpansion
\mathbb{Z}
%EndExpansion
_{2}$ topological invariant.

According to above analysis, we can get two boardly suitable conclusions
from the general form of $H_{BdG}\left( \boldsymbol{k}\right) $ (\ref{G_BdG}%
):

1. Once $\Delta ^{\ast }\left( \boldsymbol{k}\right) =\Delta \left( -%
\boldsymbol{k}\right) $, which is the situation for most cases, we have $%
H_{\sigma BdG}^{\ast }\left( \boldsymbol{k}\right) =H_{\sigma BdG}\left( -%
\boldsymbol{k}\right) $ if $h^{\ast }\left( \boldsymbol{k}\right) =h\left( -%
\boldsymbol{k}\right) $. It means $T_{+}=\mathit{I}_{2l}$ is the identity
matrix and leads to $H_{\sigma BdG}$ in the class $\mathcal{AI}$ with $\eta
_{+}$, a trivial topological classification. So only $h^{\ast }\left( 
\boldsymbol{k}\right) \neq h\left( -\boldsymbol{k}\right) $ results in a
non-trivial class and may correspond to a non-trivial topology.

2. If $\Delta \left( \boldsymbol{k}\right) $ is a zero matrix, i.e., there
is no coupling between $\boldsymbol{\beta }_{\boldsymbol{k}}^{\dagger }$ and 
$\boldsymbol{\beta }_{-\boldsymbol{k}}^{\dagger }$,%
\begin{equation}
H_{\sigma BdG}\left( \boldsymbol{k}\right) =\left( 
\begin{array}{cc}
h\left( \boldsymbol{k}\right)  & 0 \\ 
0 & -h^{\ast }\left( -\boldsymbol{k}\right) 
\end{array}%
\right) 
\end{equation}%
is still Hermitian, the topological phase of the system can be easily
distinguished by the conventional topological classification in Hermitian
systems \cite{Sch}. In the rest of the paper, we use capital letter as A to
express the topological class of Hermitian systems, while using calligraphic
letter as $\mathcal{A}$ to show the one of non-Hermitian systems. Moreover,
all physics of the system can be described by the BdG Hamiltonian $h\left( 
\boldsymbol{k}\right) $, instead of $H_{\sigma BdG}\left( \boldsymbol{k}%
\right) $. In this condition the $-h^{\ast }\left( -\boldsymbol{k}\right) $
part has become a identical copy of $h\left( \boldsymbol{k}\right) $, which
is redundant and unphysical as the BdG Hamiltonian for fermionic systems 
\cite{Zir}. Then we only need to focus on $h$, which is just the situation
that happens in the magnon systems of ferromagnets on honeycomb lattices.
The detailed information of topological classification for magnonic
Hamiltonians in monolayer and bilayer ferromagnets on honeycomb lattices is
summarized in the Table \ref{Table I}. In the following, we will study these
classification and related edge states in the specific cases.

\begin{table*}[tbph]
\caption{Topological classification for different effective 2D bulk magnonic
Hamiltonians $h(\boldsymbol{k})$ of monolayer and bilayer ferromagnets on
honeycomb lattices. The related Chern number $C_{n}$ and number of edge
states $N_{ES}$ are also shown. The topological classification is based on
the presence or absence ($0$) of time-reversal ($T$), particle-hole ($C$),
and chiral ($S$) symmetries \protect\cite{Sch}, where all three symmetric
operators are unitary, i.e., $O^{\dagger }O=1,O=T,C,S$ . They satisfy $%
T^{\dagger }h^{\ast }(\boldsymbol{k})T=h(-\boldsymbol{k})$, $C^{\dagger
}h^{\ast }(\boldsymbol{k})C=-h(-\boldsymbol{k})$, $S^{\dagger }h(\boldsymbol{%
k})S=-h(\boldsymbol{k})$, respectively. $\pm $ in $T$ and $C$ comes from $%
T^{\ast }T=\pm 1$ and $C^{\ast }C=\pm 1$. If there is AFM interaction, such
as in Fig. \protect\ref{fig1}, The topological classification is based on
the presence or absence ($0$) of $T$ and pesudo-Hermiticity $\protect\eta $ 
\protect\cite{Sat}, as shown in Sec. \protect\ref{2}D. $\mathit{I}_{N}$ is $%
N\times N$ identity matrix. $\protect\sigma _{\protect\alpha =x,y,z}$
represents Pauli matrices. $h^{\mathrm{bea}}(\boldsymbol{k})$ [$h_{\mathrm{FM%
}}^{\mathrm{bea-bea}}(\boldsymbol{k})$] means the monolayer (bilayer) case
on the honeycomb lattices with bearded boundary in Fig. \protect\ref{fig2}%
(a), $\mathrm{zig}$ ($\mathrm{arm}$) means the one with zigzag (armchair)
boundary in Fig. \protect\ref{fig2}(b) [Fig. \protect\ref{fig2}(c)],
respectively. Here, the lower index $\mathrm{FM/AFM}$ means the type of
interlayer interaction in bilayer cases. $AA/AB$ represents the stacking
form of the bilayer ferromagnets.}
\label{Table I}
\begin{center}
\setlength{\arrayrulewidth}{0.5mm} %adjust width of lines
\renewcommand\tabcolsep{5.5pt} %adjust width between nearest column
%\renewcommand\arraystretch{1.5}
%\begin{tabular}{p{3cm} p{5cm} p{1cm} p{1cm} p{1cm} p{1.5cm} p{1cm} p{1cm}}
\begin{tabular}{cccccccc}
\hline\hline
&  &  &  &  &  &  &  \\[-1ex] 
Effective 2D bulk Hamiltonian & $T$ & $C$ & $S$ & $\eta $ & Class & $C_{n}$
& $N_{ES}$ \\[0.5ex] \hline
&  &  &  &  &  &  &  \\[-1ex] 
$h^{\mathrm{bea/zig}}(\boldsymbol{k}),D=0$ [Eq. (\ref{m_bea})] & $\mathit{I}%
_{2}(+)$ & $\sigma _{z}(+)$ & $\sigma _{z}$ & $0$ & $\mathrm{BDI}$ & $0$ & $%
4/2$ (Fig. \ref{fig3}) \\[0.5ex] 
$h^{\mathrm{bea/zig}}(\boldsymbol{k}),D\neq 0$ [Eq. (\ref{m_bea})] & $0$ & $%
\sigma _{z}$ & $0$ & $0$ & $\mathrm{D}$ & $\pm 1$ & $4$ (Fig. \ref{fig3}) \\%
[0.5ex] 
$h^{\mathrm{arm}}(\boldsymbol{k}),D=0$ [(Eq. (\ref{m_arm})] & $\mathit{I}%
_{4}(+)$ & $\mathit{I}_{2}\otimes \sigma _{z}(+)$ & $\mathit{I}_{2}\otimes
\sigma _{z}$ & $0$ & $\mathrm{BDI}$ & $0$ & $4$ [Fig. \ref{fig4}(b)] \\%
[0.5ex] 
$h^{\mathrm{arm}}(\boldsymbol{k}),D\neq 0$ [Eq. (\ref{m_arm})] & $0$ & $%
\mathit{I}_{2}\otimes \sigma _{z}$ & $0$ & $0$ & $\mathrm{D}$ & $\pm 1$ & $4$
[Fig. \ref{fig4}(b)] \\[0.5ex] 
$h_{\mathrm{FM}}^{\mathrm{bea-bea/zig-zig}}(\boldsymbol{k}),D=0,AA$ [Eq. (%
\ref{m_bea})] & $\mathit{I}_{2}(+)$ & $\sigma _{z}(+)$ & $\sigma _{z}$ & $0$
& $\mathrm{BDI}$ & $0$ & $8/4$ (Fig. \ref{fig3}) \\[0.5ex] 
$h_{\mathrm{FM}}^{\mathrm{bea-bea/zig-zig}}(\boldsymbol{k}),D\neq 0,AA$ [Eq.
(\ref{m_bea})] & $0$ & $\sigma _{z}(+)$ & $0$ & $0$ & $\mathrm{D}$ & $\pm 1$
& $8$ (Fig. \ref{fig3}) \\[0.5ex] 
$h_{\mathrm{FM}}^{\mathrm{arm-arm}}(\boldsymbol{k}),D=0,AA$ [Eq. (\ref{m_arm}%
)] & $\mathit{I}_{4}(+)$ & $\mathit{I}_{2}\otimes \sigma _{z}(+)$ & $\mathit{%
I}_{2}\otimes \sigma _{z}$ & $0$ & $\mathrm{BDI}$ & $0$ & $8$ [Fig. \ref%
{fig4}(b)] \\[0.5ex] 
$h_{\mathrm{FM}}^{\mathrm{arm-arm}}(\boldsymbol{k}),D\neq 0,AA$ [Eq. (\ref%
{m_arm})] & $0$ & $\mathit{I}_{2}\otimes \sigma _{z}$ & $0$ & $0$ & $\mathrm{%
D}$ & $\pm 1$ & $8$ [Fig. \ref{fig4}(b)] \\[0.5ex] 
$\eta h_{\mathrm{AFM}}^{\mathrm{bea-bea/zig-zig}}(\boldsymbol{k}),D=0,AA$
[Eq. (\ref{AFM_bea})] & $\mathit{I}_{4}(+)$ & $0$ & $0$ & $\sigma
_{z}\otimes \mathit{I}_{2}$ & $\mathcal{AI},\eta _{+}$ & $0$ & $8/4$ (Fig. %
\ref{fig3}) \\[0.5ex] 
$\eta h_{\mathrm{AFM}}^{\mathrm{bea-bea/zig-zig}}(\boldsymbol{k}),D\neq 0,AA$
[Eq. (\ref{AFM_bea})] & $0$ & $0$ & $0$ & $\sigma _{z}\otimes \mathit{I}_{2}$
& $\mathcal{A}$ & $\pm 1$ & $8$ (Fig. \ref{fig3}) \\[0.5ex] 
$h_{\mathrm{FM}}^{\mathrm{bea-bea}}(\boldsymbol{k}),D=0,AB$ [Eq. (\ref%
{FM_AB_bea})] & $\mathit{I}_{4}(+)$ & $0$ & $0$ & $0$ & $\mathrm{AI}$ & $0$
& $8$ [Fig. \ref{fig9}(c)] \\[0.5ex] 
$h_{\mathrm{FM}}^{\mathrm{bea-bea}}(\boldsymbol{k}),D\neq 0,AB$ [Eq. (\ref%
{FM_AB_bea})] & $0$ & $0$ & $0$ & $0$ & $\mathrm{A}$ & $\pm 2$ & $8$ [Fig. %
\ref{fig9}(c)] \\[0.5ex] 
$\eta h_{\mathrm{AFM}}^{\mathrm{bea-bea}}(\boldsymbol{k}),D=0,AB$ [Eq. (\ref%
{AFM_AB_bea})] & $\mathit{I}_{4}(+)$ & $0$ & $0$ & $\sigma _{z}\otimes 
\mathit{I}_{2}$ & $\mathcal{AI,\eta _{+}}$ & $0$ & $8$ [Fig. \ref{fig10}(c)]
\\[0.5ex] 
$\eta h_{\mathrm{AFM}}^{\mathrm{bea-bea}}(\boldsymbol{k}),D\neq 0,AB$ [Eq. (%
\ref{AFM_AB_bea})] & $0$ & $0$ & $0$ & $\sigma _{z}\otimes \mathit{I}_{2}$ & 
$\mathcal{A}$ & $\pm 2$ & $8$ [Fig. \ref{fig10}(c)] \\[0.5ex] \hline
\end{tabular}%
\end{center}
\end{table*}

\begin{figure}[tbp]
\begin{center}
\includegraphics[width=0.48\textwidth]{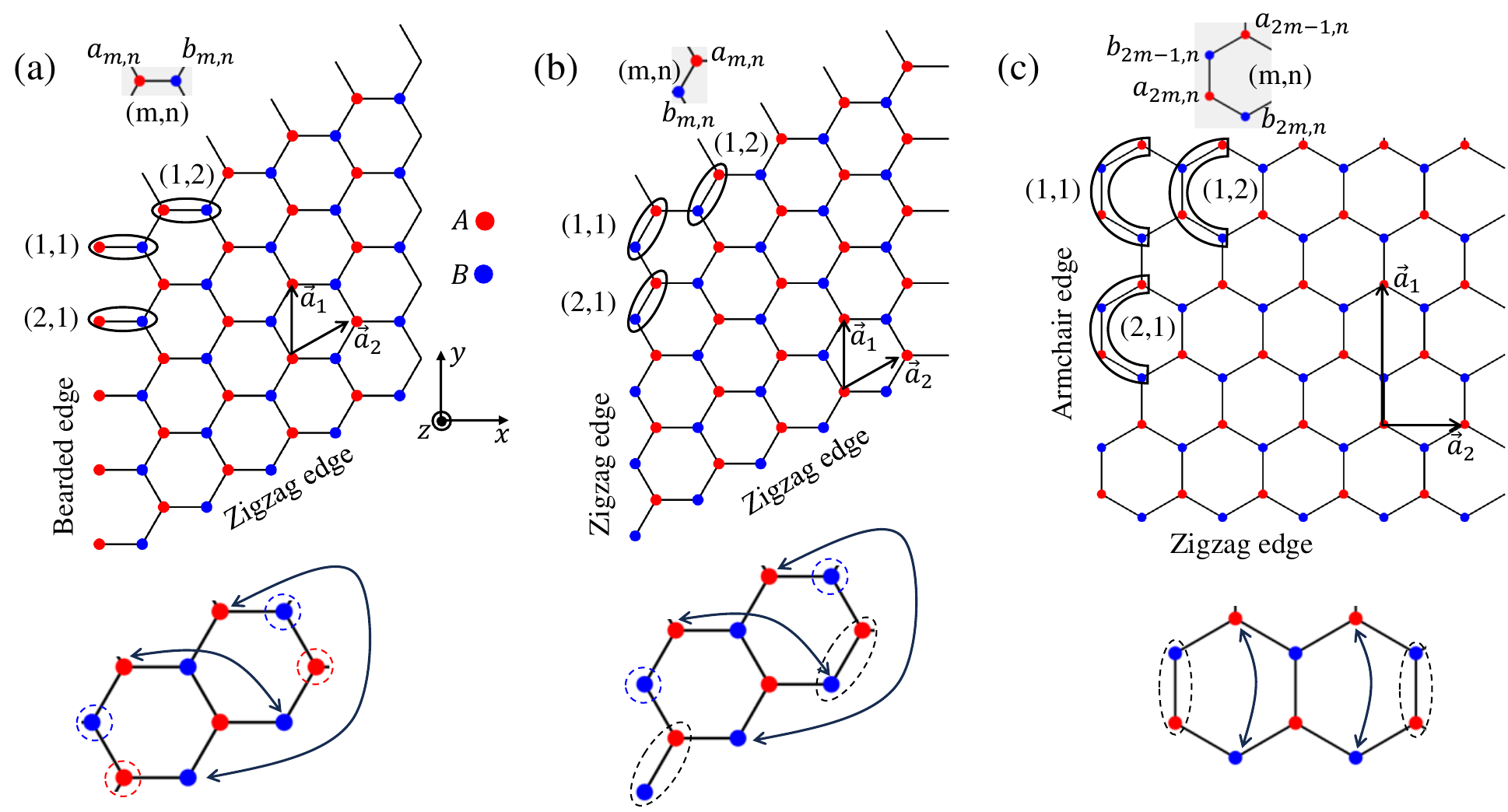}
\end{center}
\caption{(Color online) Upper panel: The schematic illustration of bosonic
lattice models for magnons in monolayer Ferromagnets on honeycomb lattices
with different edges. The primitive vectors are $\vec{a}_{1}$ and $\vec{a}%
_{2}$. The unit with translational symmetry in the Heisenberg Hamiltonians
is emphasized by the black box in each structure, respectively. The number
pairs $(1,1)$, $(1,2)$, $(2,1)$ in each figure indicate the increasing
direction of $m$ and $n$. Lower panel: The bosonic lattice structure with
the periodic boundary condition (PBC) for $N=2,M=2$ in both (a) and (b), and 
$N=2,M=1$ in (c), respectively. The lattice sites highlighted by the dashed
circle in the same color means the equivalent lattice sites due to the PBC.
The black double-head arrow represents the coupling arises from the PBC in
the $y$ direction.}
\label{fig2}
\end{figure}

\section{Topological classification and related edge states for magnons in
monolayer ferromagnets on honeycomb lattices}

\label{3}

\subsection{Bulk BdG Hamiltonian}

According to above analysis, the BdG Hamiltonian after the HP transformation
for monolayer ferromagnets on honeycomb lattices is%
\begin{eqnarray}
H_{\mathrm{FM}} &=&-3MNJS^{2}  \label{mb} \\
&&+SJ\sum_{m=1}^{M}\sum_{n=1}^{N}[3\left( a_{m,n}^{\dagger
}a_{m,n}+b_{m,n}^{\dagger }b_{m,n}\right)  \notag \\
&&-(a_{m,n}^{\dagger }b_{m,n}+b_{m,n}^{\dagger }a_{m,n+1}  \notag \\
&&+b_{m,n}^{\dagger }a_{m+1,n+1}+H.c.)]  \notag \\
&&+H_{\mathrm{DM}}+H_{\mathrm{ON}}+H_{\mathrm{Z}},  \notag
\end{eqnarray}%
with%
\begin{eqnarray*}
H_{\mathrm{DM}} &=&SDi\sum_{m=1}^{M}\sum_{n=1}^{N}(a_{m,n}^{\dagger
}a_{m,n+1}+a_{m,n}^{\dagger }a_{m+1,n} \\
&&-a_{m,n}^{\dagger }a_{m+1,n+1})+\mathrm{H.c.}-\left( a\longleftrightarrow
b\right) ,
\end{eqnarray*}%
\begin{equation*}
H_{\mathrm{ON}}=MN\mathcal{K}S^{2}-2S\mathcal{K}\sum_{m=1}^{M}\sum_{n=1}^{N}%
\left( a_{m,n}^{\dagger }a_{m,n}+b_{m,n}^{\dagger }b_{m,n}\right) ,
\end{equation*}%
\begin{equation*}
H_{\mathrm{Z}}=\frac{g\mu _{B}}{\hbar }B_{z}\left[ MNS-\sum_{m=1}^{M}%
\sum_{n=1}^{N}\left( a_{m,n}^{\dagger }a_{m,n}+b_{m,n}^{\dagger
}b_{m,n}\right) \right] ,
\end{equation*}%
which comes from the assumption that the system is infinite (periodic) along
both $\vec{a}_{1}$ and $\vec{a}_{2}$ directions in Fig. \ref{fig2}(a). Here $%
H_{\mathrm{DM}}$ represents the NNN DMI, $H_{\mathrm{ON}}$ describes the
easy-axis anisotropic term, and $H_{\mathrm{Z}}$ means the Zeeman term.

In the rest of the paper, we ignore the constant term such as $-3MNJS^{2}$
in Eq. (\ref{mb}) for simplicity since it cannot influence the properties of
the system. Taking the Fourier transformation%
\begin{equation}
f_{\boldsymbol{k}}=\sum_{m,n=1}^{M,N}e^{-i\left( nk_{x}+mk_{y}\right)
}f_{m,n},f=a,b,
\end{equation}%
the Hamiltonian in $\boldsymbol{k}$ space can be written as%
\begin{eqnarray}
H\left( \boldsymbol{k}\right)  &=&\frac{S}{2}J\{3\sum_{f=a,b}f_{\boldsymbol{k%
}}^{\dagger }f_{\boldsymbol{k}}-\left[ 1+e^{-ik_{x}}+e^{-i\left(
k_{x}+k_{y}\right) }\right] a_{\boldsymbol{k}}^{\dagger }b_{\boldsymbol{k}} 
\notag \\
&&+\mathrm{H.c.}\}+\frac{1}{2}\left[ H_{\mathrm{DM}}\left( \boldsymbol{k}%
\right) +H_{\mathrm{ON}}\left( \boldsymbol{k}\right) +H_{\mathrm{Z}}\left( 
\boldsymbol{k}\right) \right]   \notag \\
&&+\left( \boldsymbol{k}\leftrightarrow -\boldsymbol{k}\right)   \notag \\
&=&\eta _{k}^{\dagger }H_{BdG}\left( \boldsymbol{k}\right) \eta _{k},\eta
_{k}=\left( a_{\boldsymbol{k}},b_{\boldsymbol{k}},a_{-\boldsymbol{k}%
}^{\dagger },b_{-\boldsymbol{k}}^{\dagger }\right) ^{T},
\end{eqnarray}%
with%
\begin{eqnarray*}
H_{\mathrm{DM}}\left( \boldsymbol{k}\right)  &=&2SD\left[ \sin \left(
k_{x}+k_{y}\right) -\sin k_{x}-\sin k_{y}\right]  \\
&&\times \left( a_{\boldsymbol{k}}^{\dagger }a_{\boldsymbol{k}}-b_{%
\boldsymbol{k}}^{\dagger }b_{\boldsymbol{k}}\right) ,
\end{eqnarray*}%
\begin{eqnarray*}
H_{\mathrm{ON}}\left( \boldsymbol{k}\right)  &=&-2S\mathcal{K}\sum_{f=a,b}f_{%
\boldsymbol{k}}^{\dagger }f_{\boldsymbol{k}}, \\
H_{\mathrm{Z}}\left( \boldsymbol{k}\right)  &=&-\frac{g\mu _{B}}{\hbar }%
B_{z}\sum_{f=a,b}f_{\boldsymbol{k}}^{\dagger }f_{\boldsymbol{k}}.
\end{eqnarray*}%
This causes%
\begin{eqnarray}
H_{BdG}^{\mathrm{bea}}\left( \boldsymbol{k}\right)  &=&\left( 
\begin{array}{cc}
h\left( \boldsymbol{k}\right)  & 0 \\ 
0 & h^{\ast }\left( -\boldsymbol{k}\right) 
\end{array}%
\right) ,  \label{m_bea} \\
h\left( \boldsymbol{k}\right)  &=&\frac{S}{2}\left[ \gamma \mathit{I}%
_{2}+\left( 
\begin{array}{cc}
\beta  & -\alpha  \\ 
-\alpha ^{\ast } & -\beta 
\end{array}%
\right) \right] ,  \notag
\end{eqnarray}%
where%
\begin{eqnarray}
\alpha  &=&J\left[ 1+e^{-ik_{x}}+e^{-i\left( k_{x}+k_{y}\right) }\right] , 
\notag \\
\beta  &=&2D\left[ \sin \left( k_{x}+k_{y}\right) -\sin k_{x}-\sin k_{y}%
\right] ,  \notag \\
\gamma  &=&3J-2\mathcal{K}-\frac{g\mu _{B}}{S\hbar }B_{z},  \notag
\end{eqnarray}%
bea means the bearded boundary [Fig. \ref{fig2}(a)] we considered here. The
straightforward calculation shows the eigenenergies and eigenstates of $%
h\left( \boldsymbol{k}\right) $ can be expressed as 
\begin{equation}
\varepsilon _{\pm }=\frac{S}{2}\left( \gamma \pm \sqrt{\beta ^{2}+\left\vert
\alpha \right\vert ^{2}}\right) =\frac{S}{2}\left( \gamma \pm \varepsilon
\right) ,
\end{equation}%
\begin{eqnarray*}
\left\vert u_{+}\right\rangle  &=&\frac{1}{\sqrt{\Omega }}\left\{ 
\begin{array}{c}
\beta +\varepsilon  \\ 
-\alpha ^{\ast }%
\end{array}%
\right\} , \\
\left\vert u_{-}\right\rangle  &=&\frac{1}{\sqrt{\Omega }}\left\{ 
\begin{array}{c}
\alpha  \\ 
\beta +\varepsilon 
\end{array}%
\right\} ,
\end{eqnarray*}%
with $\Omega =2\varepsilon \left( \varepsilon +\beta \right) $ is the
normalization factor. Combining with the symmetry analysis in Table \ref%
{Table I}, the Berry curvature of these two magnon bands can be calculated as%
\begin{equation*}
\Omega _{n}^{k_{x}k_{y}}=i\sum_{l\neq n}\left( \mathcal{R}_{n,l}^{k_{x}}%
\mathcal{R}_{l,n}^{k_{y}}-\mathcal{R}_{n,l}^{k_{y}}\mathcal{R}%
_{l,n}^{k_{x}}\right) ,
\end{equation*}%
where the Berry connection $\mathcal{R}_{n,l}^{k_{\alpha }}\equiv
\left\langle u_{n}\right\vert \partial _{k_{\alpha }}\left\vert
u_{l}\right\rangle $, $\alpha =x,y$ and $n,l=\pm $. One can easily find that%
\begin{eqnarray}
C_{-} &=&-C_{+}=\frac{1}{2\pi }\int_{BZ^{d}}\Omega _{-}^{k_{x}k_{y}}\mathrm{d%
}\boldsymbol{k}  \notag \\
&\boldsymbol{=}&\left\{ 
\begin{array}{c}
1,D<0 \\ 
0,D=0 \\ 
-1,D>0%
\end{array}%
\right. ,
\end{eqnarray}%
which is a $%
%TCIMACRO{\U{2124} }%
%BeginExpansion
\mathbb{Z}
%EndExpansion
$ topological invariant and independent with $J$. For the BdG Hamiltonian $%
H_{BdG}^{\mathrm{zig}}\left( \boldsymbol{k}\right) $ represented in the
bosonic honeycomb lattice shown in Fig. \ref{fig2}(b), the same bulk
Hamiltonian can be found after the gauge transformation. The equivalence of
the lattice structures under PBC shown in lower panel of Figs. \ref{fig2}(a)
and (b) also exhibits this point.

However, for the situation in Fig. \ref{fig2}(c), i.e., the bosonic
honeycomb lattice with the armchair boundary, we have%
\begin{eqnarray}
H_{\mathrm{FM}} &=&SJ\sum_{m=1}^{M}\sum_{n=1}^{N}[3(a_{2m-1,n}^{\dagger
}a_{2m-1,n}  \notag \\
&&+b_{2m-1,n}^{\dagger }b_{2m-1,n}+a_{2m,n}^{\dagger }a_{2m,n}  \notag \\
&&+b_{2m,n}^{\dagger }b_{2m,n})-b_{2m-1,n}^{\dagger }\left(
a_{2m-1,n}+a_{2m,n}\right)   \notag \\
&&-b_{2m,n}^{\dagger }\left( a_{2m,n}+a_{2m+1,n}\right)   \notag \\
&&-a_{2m-1,n}^{\dagger }b_{2m-1,n+1}-b_{2m,n}^{\dagger }a_{2m,n+1}  \notag \\
&&+H.c.]+H_{\mathrm{DM}}+H_{\mathrm{ON}}+H_{\mathrm{Z}},
\end{eqnarray}%
\begin{eqnarray*}
H_{\mathrm{DM}} &=&SDi\sum_{m=1}^{M}\sum_{n=1}^{N}[a_{2m-1,n}^{\dagger
}a_{2m,n}-a_{2m,n}^{\dagger }a_{2m+1,n} \\
&&+a_{2m-1,n}^{\dagger }a_{2m-1,n+1}+a_{2m,n}^{\dagger }a_{2m,n+1} \\
&&-a_{2m-1,n}^{\dagger }a_{2m,n+1}+b_{2m-1,n}^{\dagger }b_{2m,n} \\
&&-b_{2m,n}^{\dagger }b_{2m+1,n}-(b_{2m-1,n}^{\dagger }b_{2m-1,n+1} \\
&&+b_{2m,n}^{\dagger }b_{2m,n+1}-b_{2m,n}^{\dagger }b_{2m-1,n+1})+H.c.],
\end{eqnarray*}%
since the minimum repetitive unit includes four sites instead of two sites.
Still ignoring the constant term, the Fourier transformation gives%
\begin{eqnarray}
H\left( \boldsymbol{k}\right)  &=&\frac{S}{2}J[3\sum_{f=a,b}\left( f_{1,%
\boldsymbol{k}}^{\dagger }f_{1,\boldsymbol{k}}+f_{2,\boldsymbol{k}}^{\dagger
}f_{2,\boldsymbol{k}}\right)  \\
&&-\left( 1+e^{ik_{x}}\right) a_{1,\boldsymbol{k}}^{\dagger }b_{1,%
\boldsymbol{k}}  \notag \\
&&-\left( 1+e^{-ik_{x}}\right) a_{2,\boldsymbol{k}}^{\dagger }b_{2,%
\boldsymbol{k}}-a_{2,\boldsymbol{k}}^{\dagger }b_{1,\boldsymbol{k}}  \notag
\\
&&-e^{-ik_{y}}a_{1,\boldsymbol{k}}^{\dagger }b_{2,\boldsymbol{k}}+H.c.] 
\notag \\
&&+\frac{1}{2}\left[ H_{\mathrm{DM}}\left( \boldsymbol{k}\right) +H_{\mathrm{%
ON}}\left( \boldsymbol{k}\right) +H_{\mathrm{Z}}\left( \boldsymbol{k}\right) %
\right]   \notag \\
&&+\left( \boldsymbol{k}\leftrightarrow -\boldsymbol{k}\right)   \notag \\
&=&\eta _{k}^{\dagger }H_{BdG}\left( \boldsymbol{k}\right) \eta _{k},\eta
=\left( \xi _{\boldsymbol{k}},\xi _{-\boldsymbol{k}}^{\dagger }\right) ^{T},
\notag
\end{eqnarray}%
with $\xi _{\boldsymbol{k}}=\left( a_{1,\boldsymbol{k}},b_{1,\boldsymbol{k}%
},a_{2,\boldsymbol{k}},b_{2,\boldsymbol{k}}\right) $,%
\begin{eqnarray*}
H_{\mathrm{DM}}\left( \boldsymbol{k}\right)  &=&-2SD\sin k_{x}\left( a_{1,%
\boldsymbol{k}}^{\dagger }a_{1,\boldsymbol{k}}+a_{2,\boldsymbol{k}}^{\dagger
}a_{2,\boldsymbol{k}}\right)  \\
&&+SDi\left[ \left( 1+e^{-ik_{y}}-e^{ik_{x}}\right) a_{1,\boldsymbol{k}%
}^{\dagger }a_{2,\boldsymbol{k}}\right] +H.c. \\
&&+2SD\sin k_{x}\left( b_{1,\boldsymbol{k}}^{\dagger }b_{1,\boldsymbol{k}%
}+b_{2,\boldsymbol{k}}^{\dagger }b_{2,\boldsymbol{k}}\right)  \\
&&+SDi\left[ \left( 1+e^{-ik_{y}}-e^{-ik_{x}}\right) b_{1,\boldsymbol{k}%
}^{\dagger }b_{2,\boldsymbol{k}}\right] +H.c.,
\end{eqnarray*}%
\begin{eqnarray*}
H_{\mathrm{ON}}\left( \boldsymbol{k}\right)  &=&-2S\mathcal{K}%
\sum_{f=a,b}\left( f_{1,\boldsymbol{k}}^{\dagger }f_{1,\boldsymbol{k}}+f_{2,%
\boldsymbol{k}}^{\dagger }f_{2,\boldsymbol{k}}\right) , \\
H_{\mathrm{Z}}\left( \boldsymbol{k}\right)  &=&-\frac{g\mu _{B}}{\hbar }%
B_{z}\sum_{f=a,b}\left( f_{1,\boldsymbol{k}}^{\dagger }f_{1,\boldsymbol{k}%
}+f_{2,\boldsymbol{k}}^{\dagger }f_{2,\boldsymbol{k}}\right) ,
\end{eqnarray*}%
leading to%
\begin{eqnarray}
H_{BdG}^{\mathrm{arm}}\left( \boldsymbol{k}\right)  &=&\left( 
\begin{array}{cc}
h\left( \boldsymbol{k}\right)  & 0 \\ 
0 & h^{\ast }\left( -\boldsymbol{k}\right) 
\end{array}%
\right) ,  \label{m_arm} \\
h\left( \boldsymbol{k}\right)  &=&\frac{S}{2}\left[ \gamma \mathit{I}%
_{4}+\left( 
\begin{array}{cccc}
-\beta  & -\alpha  & \lambda _{1} & -Je^{-ik_{y}} \\ 
-\alpha ^{\ast } & \beta  & -J & \lambda _{2} \\ 
\lambda _{1}^{\ast } & -J & -\beta  & -\alpha ^{\ast } \\ 
-Je^{ik_{y}} & \lambda _{2}^{\ast } & -\alpha  & \beta 
\end{array}%
\right) \right] ,  \notag
\end{eqnarray}%
with%
\begin{eqnarray}
\alpha  &=&J\left( 1+e^{ik_{x}}\right) ,\beta =2D\sin k_{x},  \notag \\
\lambda _{1} &=&Di\left[ \left( 1+e^{-ik_{y}}-e^{ik_{x}}\right) \right] , 
\notag \\
\lambda _{2} &=&Di\left[ \left( 1+e^{-ik_{y}}-e^{-ik_{x}}\right) \right] , 
\notag \\
\gamma  &=&3J-2\mathcal{K}-\frac{g\mu _{B}}{S\hbar }B_{z}.
\end{eqnarray}%
From the lattice structure under PBC shown in lower panel of Fig. \ref{fig2}%
(c), it is obvious that the bulk Hamiltonian with the armchair boundary is
different from the one with bearded or zigzag boundary due to the PBC in the 
$y$ direction. When $D=0$, the energy dispersions have forms%
\begin{eqnarray*}
\varepsilon _{1,2} &=&\frac{S}{2}\left( \gamma -\varepsilon _{\pm }\right)
,\varepsilon _{3,4}=\frac{S}{2}\left( \gamma +\varepsilon _{\pm }\right) , \\
\varepsilon _{\pm } &=&\sqrt{\left( 2\cos \frac{k_{x}}{2}\pm \cos \frac{k_{y}%
}{2}\right) ^{2}+\sin ^{2}\frac{k_{y}}{2}}.
\end{eqnarray*}%
For $D\neq 0$, the straightforward numerical calculation shows $\varepsilon
_{1}$ and $\varepsilon _{2}$ ($\varepsilon _{3}$ and $\varepsilon _{4}$) are
gapless, but there is always a gap between $\varepsilon _{2}$ and $%
\varepsilon _{3}$. In this situation, although the Chern number for a single
band is no longer well-defined due to the degeneracy, the sum of Chern
numbers of different bands can still be effective, where our numerical
calculation exhibits that%
\begin{eqnarray*}
C_{1}+C_{2} &=&-\left( C_{3}+C_{4}\right)  \\
&=&\frac{1}{2\pi }\int_{BZ^{d}}\left( \Omega _{1}^{k_{x}k_{y}}+\Omega
_{2}^{k_{x}k_{y}}\right) \mathrm{d}\boldsymbol{k} \\
&=&\left\{ 
\begin{array}{c}
1,D<0 \\ 
-1,D>0%
\end{array}%
\right. ,
\end{eqnarray*}%
with%
\begin{equation*}
\Omega _{1}^{k_{x}k_{y}}+\Omega
_{2}^{k_{x}k_{y}}=i\sum_{n=1,2}\sum_{l=3,4}\left( \mathcal{R}_{n,l}^{k_{x}}%
\mathcal{R}_{l,n}^{k_{y}}-\mathcal{R}_{n,l}^{k_{y}}\mathcal{R}%
_{l,n}^{k_{x}}\right) ,
\end{equation*}%
which only corresponds with the Berry connections between bands $1,2$ and $%
3,4$.

\begin{figure*}[tbp]
\begin{center}
\includegraphics[width=0.75\textwidth]{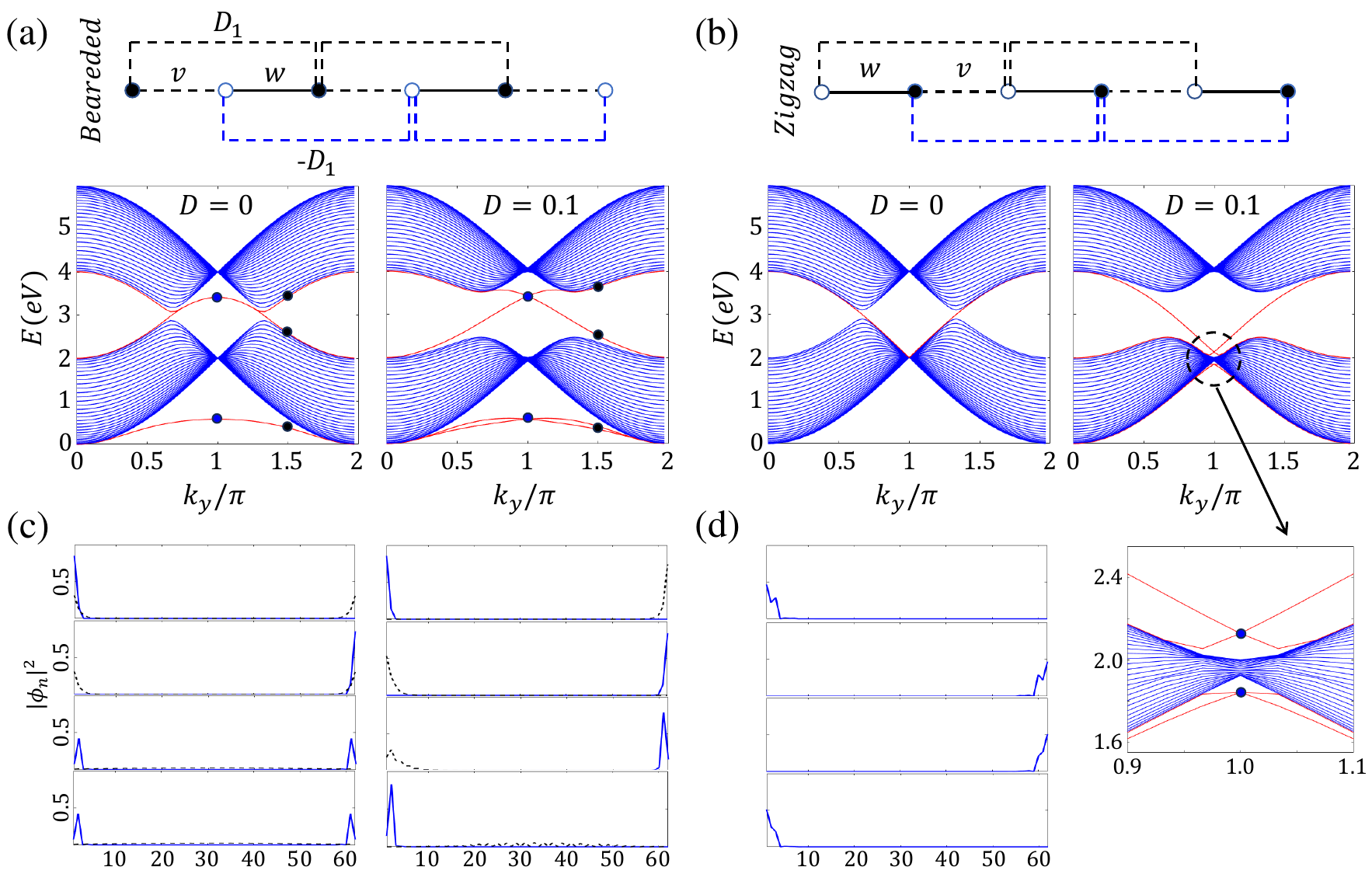}
\end{center}
\caption{(Color online) (a) Top panel: The schematic illustration of 1D
effective SSH chain parameterized by $k_{y}$ for magnons in monolayer
ferromagnets on honeycomb lattices with the bearded boundary shown in Fig. 
\protect\ref{fig2}(a). The on-site potentials are omitted for simplicity
with $w=-2SJ\cos (k_{y}/2)$, $v=-SJ$, and $D_{1}=2SD\sin (k_{y}/2)$. Bottom
panel: The band structure of this effective SSH chain for different $D$
values (in the unit of $J$), where the edge states are highlighted by the
red solid lines. (b) The similar plot for the one with the zigzag boundary
shown in Fig. \protect\ref{fig2}(b). (c) The wave function distribution of
edge states in real space at specific values of $k_{y}$ marked in (a). From
the top plot to the bottom plot corresponds with the edge state with the
lowest energy to the one with the highest energy, respectively. The black
dashed (blue soild) lines represent the edge-state distribution marked by
black (blue) circles in (a). (d) Right panel: The zoom-in plot of the band
structure shown in (b) with non-zero $D$ (in the unit of $J$). Left panel:
The similar plot as (c) at specific values of $k_{y}$ marked in the left
panel. Here $S=1$, $J=1\,\mathrm{eV}$, and $\protect\gamma =3J$ ($\mathcal{K}%
=0$, $B_{z}=0$) is used for all plots, respectively.}
\label{fig3}
\end{figure*}

\begin{figure*}[tbp]
\begin{center}
\includegraphics[width=0.75\textwidth]{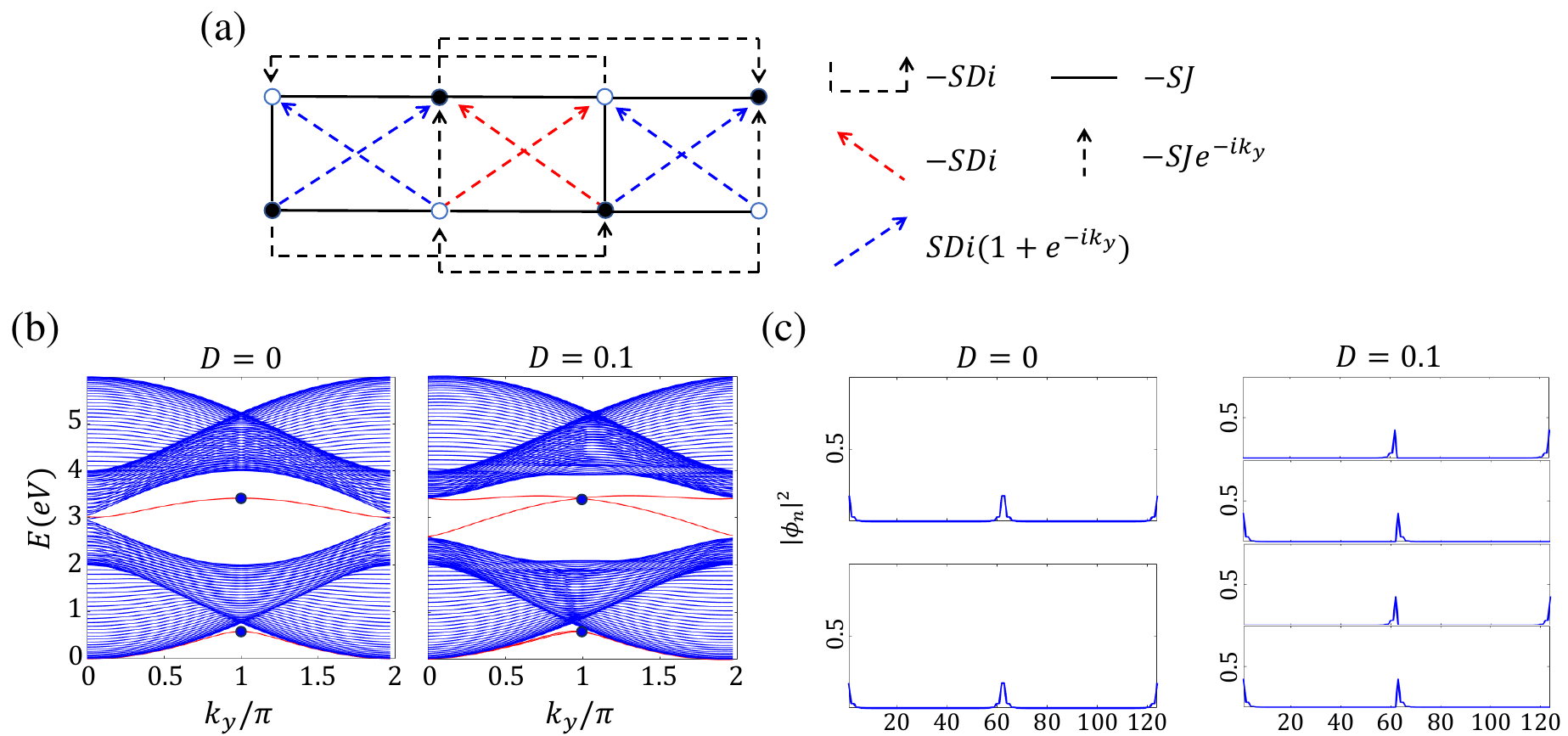}
\end{center}
\caption{(Color online) (a) Schematic illustration of quasi-1D effective
chain parameterized by $k_{y}$ for magnons in monolayer ferromagnets on
honeycomb lattices with the armchair boundary shown in Fig. \protect\ref%
{fig2}(c). The on-site potentials are omitted for simplicity, and the
coupling between different sites are denoted in the right. (b) The band
structure of this effective chain for different $D$ values (in the unit of $%
J $), where the edge states are highlighted by the red solid lines. (c) The
wave function distribution of edge states in real space at specific values
of $k_{y}$ marked in (b). From the top plot to the bottom plot corresponds
with the edge state with the lowest energy to the one with the highest
energy, respectively. There is only two plots for $D=0$ since two degenerate
edge states have the same wave function distribution. Here $S=1$, $J=1\,%
\mathrm{eV}$, and $\protect\gamma =3J$ ($\mathcal{K}=0$, $B_{z}=0$) is used
for all plots, respectively.}
\label{fig4}
\end{figure*}

\subsection{BdG Hamiltonian with open boundary conditions and related edge
states}

\begin{figure}[tbp]
\begin{center}
\includegraphics[width=0.48\textwidth]{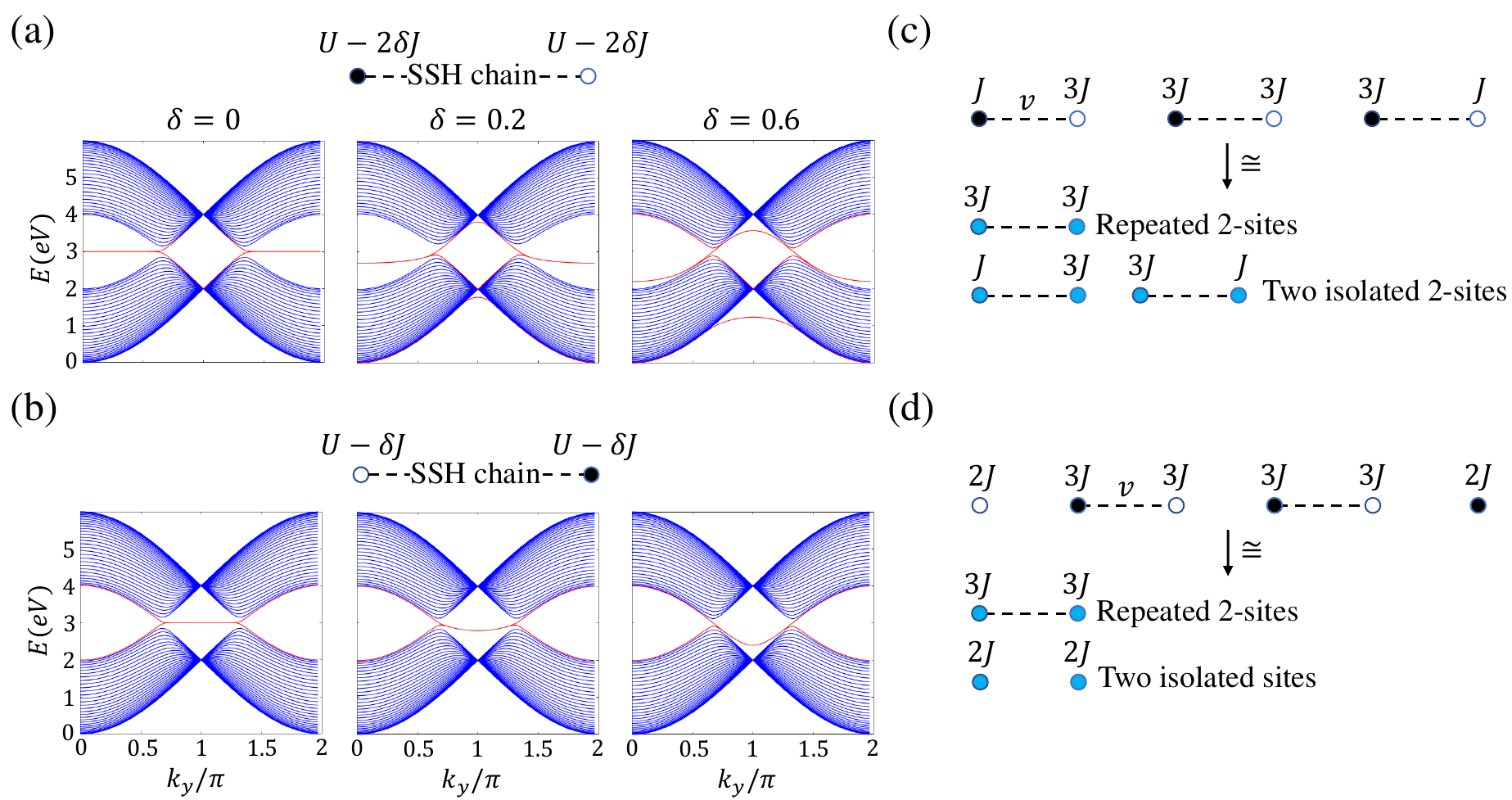}
\end{center}
\caption{(Color online) (a) The change of the band structure with the
on-site potential $U-2\protect\delta J$ at the two ends of the effective SSH
chain parametrized by $k_{y}$ in Fig. \protect\ref{fig3}(a) when $D=0$. (b)
The similar plot for the effective SSH chain in Fig. \protect\ref{fig3}(b). $%
S=1$, $J=1\,\mathrm{eV}$, and $U=3J$ have been used in both (a) and (b). (c)
The Effective SSH chain parametrized by $k_{y}$ when $k_{y}\approx \protect%
\pi $ for the bosonic honeycomb lattice shown in Fig. \protect\ref{fig2}(a).
The equivalent lattice structures are shown in blue in the figure. (d) The
similar plot as (c) for bosonic honeycomb lattice shown in Fig. \protect\ref%
{fig2}(b).}
\label{fig5}
\end{figure}

Here we mainly consider the edge states of $H_{BdG}$ based on three
conventional edges on honeycomb lattice, as shown in Fig. \ref{fig2}. By
assuming the system is infinite (periodic) along $\vec{a}_{1}$ direction and
finite along $\vec{a}_{2}$ direction, which is an usual assumption to
investigate edge states for discrete 2D systems such as graphene \cite%
{Yao,Del,Tan}, the effective 1D BdG Hamiltonians $H\left( k_{y}\right) $
parameterized by wave vector number $k_{y}$ are shown in Figs. \ref{fig3}
and \ref{fig4} for each structures. The band structure of $H^{\mathrm{%
\lambda }}\left( k_{y}\right) $, $\mathrm{\lambda }=\mathrm{bea,zig,arm}$
and related wave function distribution of edge states for specific values of 
$k_{y}$ are shown in Figs. \ref{fig3}(a), (b) and \ref{fig4}, respectively.
Interestingly, the edge states appear with the lowest or highest energy
instead of zero-energy flat bands near the Fermi level as in the fermionic
cases \cite{Yao,Del}. Taking the bosonic honeycomb lattice with the bearded
boundary [Fig. \ref{fig2}(a)] as an example, the 2D BdG Hamiltonian of this
lattice can be expressed as%
\begin{eqnarray}
H_{\mathrm{FM}} &=&SJ\sum_{m=1}^{M}\{\sum_{n=1}^{N}[3\left( a_{m,n}^{\dagger
}a_{m,n}+b_{m,n}^{\dagger }b_{m,n}\right) \\
&&-a_{m,n}^{\dagger }b_{m.n}]-2\left( a_{m,1}^{\dagger
}a_{m,1}+b_{m,N}^{\dagger }b_{m,N}\right)  \notag \\
&&-\sum_{n=1}^{N-1}\left( b_{m,n}^{\dagger }a_{m,n+1}+b_{m,n}^{\dagger
}a_{m+1,n+1}\right) +H.c.\}  \notag \\
&&+H_{\mathrm{DM}}+H_{\mathrm{ON}}+H_{\mathrm{Z}},  \notag
\end{eqnarray}%
where the term $-2SJ\sum_{m=1}^{M}\left( a_{m,1}^{\dagger
}a_{m,1}+b_{m,N}^{\dagger }b_{m,N}\right) $ coms from the boundary condition
since there is only one bond connect these sites in the boundary with the
bulk sites.%
\begin{eqnarray*}
H_{\mathrm{DM}} &=&SDi\sum_{m=1}^{M}[\sum_{n=1}^{N}a_{m,n}^{\dagger
}a_{m+1,n} \\
&&+\sum_{n=1}^{N-1}\left( a_{m,n}^{\dagger }a_{m,n+1}-a_{m,n}^{\dagger
}a_{m+1,n+1}\right) ]+H.c. \\
&&-\left( a\longleftrightarrow b\right) .
\end{eqnarray*}%
and $H_{\mathrm{ON}}$ and $H_{\mathrm{Z}}$ have the same expression as in
Eq. (\ref{mb}). This gives%
\begin{eqnarray*}
H^{\mathrm{bea}}\left( k_{y}\right) &=&S\{\sum_{n=1}^{N}[\gamma \left(
a_{k_{y},n}^{\dagger }a_{k_{y},n}+b_{k_{y},n}^{\dagger }b_{k_{y},n}\right) \\
&&-Ja_{k_{y},n}^{\dagger }b_{k_{y},n}] \\
&&-J\sum_{n=1}^{N-1}\left( 1+e^{-ik_{y}}\right) a_{k_{y},n+1}^{\dagger
}b_{k_{y},n}+H.c. \\
&&-2J\left( a_{k_{y},1}^{\dagger }a_{k_{y},1}+b_{k_{y},N}^{\dagger
}b_{k_{y},N}\right) \}+H_{\mathrm{DM}},
\end{eqnarray*}%
with $\gamma =3J-2\mathcal{K}-\frac{g\mu _{B}}{S\hbar }B_{z}$. 
\begin{eqnarray*}
H_{\mathrm{DM}} &\rightarrow &SD\sum_{n=1}^{N-1}[\left( -2\sin k_{y}\right)
a_{k_{y},n}^{\dagger }a_{k_{y},n} \\
&&+i\left( 1-e^{ik_{y}}\right) a_{k_{y},n}^{\dagger }a_{k_{y},n+1}+H.c.] \\
&&+SD\left( -2\sin k_{y}\right) a_{k_{y},N}^{\dagger }a_{k_{y},N}-\left(
a\longleftrightarrow b\right) .
\end{eqnarray*}%
By using the gauge%
\begin{equation*}
a_{k_{y},n}\rightarrow e^{-ink_{y}/2}a_{k_{y},n},b_{k_{y},n}\rightarrow
e^{-ink_{y}/2}b_{k_{y},n},
\end{equation*}%
we finally have%
\begin{eqnarray*}
H_{\mathrm{DM}} &\rightarrow &SD2\{\sum_{n=1}^{N-1}[-\sin
k_{y}a_{k_{y},n}^{\dagger }a_{k_{y},n} \\
&&+\sin \frac{k_{y}}{2}a_{k_{y},n}^{\dagger }a_{k_{y},n+1}+H.c.] \\
&&-\sin k_{y}a_{k_{y},N}^{\dagger }a_{k_{y},N}-\left( a\longleftrightarrow
b\right) \},
\end{eqnarray*}%
and%
\begin{eqnarray}
H^{\mathrm{bea}}\left( k_{y}\right) &=&S\{\sum_{n=1}^{N}[\gamma \left(
a_{k_{y},n}^{\dagger }a_{k_{y},n}+b_{k_{y},n}^{\dagger }b_{k_{y},n}\right)
\label{mbo} \\
&&-Ja_{k_{y},n}^{\dagger }b_{k_{y},n}]  \notag \\
&&-\sum_{n=1}^{N-1}2J\cos \frac{k_{y}}{2}a_{k_{y},n+1}^{\dagger
}b_{k_{y},n}+H.c.  \notag \\
&&-2J\left( a_{k_{y},1}^{\dagger }a_{k_{y},1}+b_{k_{y},N}^{\dagger
}b_{k_{y},N}\right) \}+H_{\mathrm{DM}}.  \notag
\end{eqnarray}%
Here, the edge states that appeared when $D=0$ can be well explained by the
non-trivial topological behavior of the SSH chain \cite{Tan}, where the
newly appeared edge states with the lowest energy and those with non-flat
bands within bulk bands compared with the graphene case is due to the
on-site energy in the end of the SSH chain, as shown in Fig. \ref{fig5}(a),
which can also be explained by the chain structure in Fig. \ref{fig5}(c)
when $v>>w$. For $D\neq 0$, these edge states can be connected to the Chern
number through their winding numbers \cite{Hat,Hat1,Wan} and no longer
degenerate except for $k_{y}=\pi $, which still can be briefly explained by
the similar chain structure shown in Fig. \ref{fig5}(c) since $D_{1}<<v$.

For the one with the zigzag open boundary condition [Fig. \ref{fig2}(b)],
the similar calculation gives 
\begin{eqnarray*}
H_{\mathrm{DM}} &\rightarrow &SD2\{\sum_{n=1}^{N-1}[-\sin
k_{y}a_{k_{y},n}^{\dagger }a_{k_{y},n} \\
&&+\sin \frac{k_{y}}{2}a_{k_{y},n}^{\dagger }a_{k_{y},n+1}+H.c. \\
&&-\sin k_{y}a_{k_{y},N}^{\dagger }a_{k_{y},N}]-\left( a\longleftrightarrow
b\right) \},
\end{eqnarray*}%
\begin{eqnarray*}
H^{\mathrm{zig}}\left( k_{y}\right) &=&S\{\sum_{n=1}^{N}[\gamma \left(
a_{k_{y},n}^{\dagger }a_{k_{y},n}+b_{k_{y},n}^{\dagger }b_{k_{y},n}\right) \\
&&-2J\cos \frac{k_{y}}{2}b_{k_{y},n}^{\dagger }a_{k_{y},n}] \\
&&-J\sum_{n=1}^{N-1}b_{k_{y},n+1}^{\dagger }a_{k_{y},n}+H.c. \\
&&-J\left( b_{k_{y},1}^{\dagger }b_{k_{y},1}+a_{k_{y},N}^{\dagger
}a_{k_{y},N}\right) \}+H_{\mathrm{DM}}.
\end{eqnarray*}%
In this situation, the edge states only appear within a very small area both
in $k_{y}$ and energy $E$ for $D=0$ due to the on-site energy in the end of
the SSH chain, as shown in Figs. \ref{fig3}(b), which is different from the
bearded one. For $D\neq 0$, one edge states with the in-gap band appears
again, and the other three non-degenerate edge states appear around $%
k_{y}=\pi $, as shown in Fig. \ref{fig3}(b). Similar as what we found in the
bearded case, the on-site energy in the end of the SSH chain detune the edge
states in zigzag case to disappear, as shown in Fig. \ref{fig5}(b). When $%
v>>w$, the chain structure for zigzag condition also implies the vanish of
edge states, as shown in Fig. \ref{fig5}(d).

\begin{figure}[tbp]
\begin{center}
\includegraphics[width=0.48\textwidth]{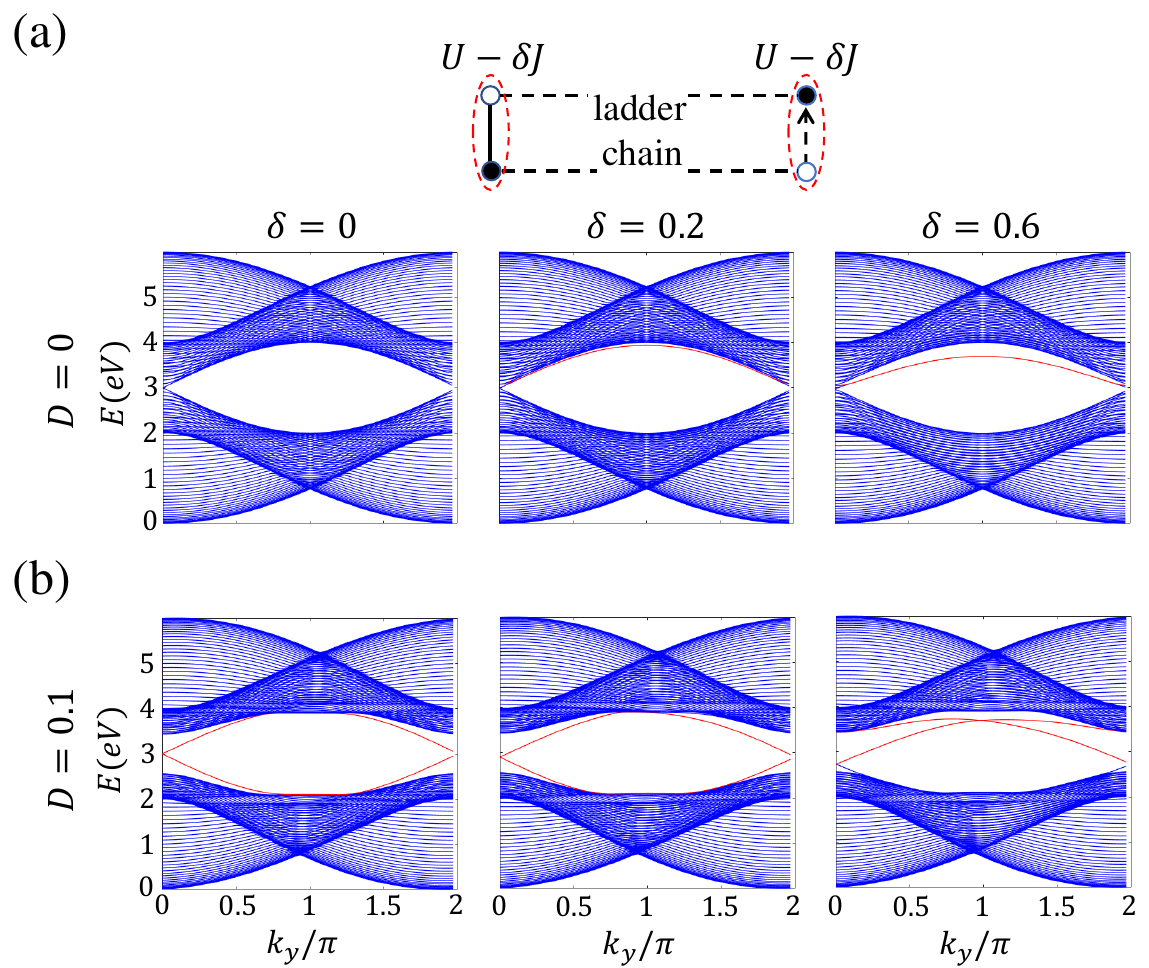}
\end{center}
\caption{(Color online) (a) The change of the band structure with the
on-site potential $U-\protect\delta J$ at the two ends of the effective
ladder chain parametrized by $k_y$ in Fig. \protect\ref{fig4}(a) when $D=0$.
(b) The similar plot for the effective ladder chain in Fig. \protect\ref%
{fig4}(a) when $D=0.1J$. $S=1$, $J=1\,\mathrm{eV}$, and $U=3J$ have been
used in both (a) and (b). }
\label{fig6}
\end{figure}

Finally, for the one with the armchair open boundary condition [Fig. \ref%
{fig2}(c)], the straightforward derivation indicates that the lattice
structure of the BdG Hamiltonian $H^{\mathrm{arm}}\left( k_{y}\right) $ is
no longer the SSH-like type. Instead, it becomes a ladder-like
quasi-one-dimensional chain, as shown in Fig. \ref{fig4}(a). The edge states
still appear for both $D=0$ and $D\neq 0$ situations. For $D=0$, the edge
states are always two-fold degenerate, which arise from the difference of
the on-site potentials in the boundary between those in the bulk, as shown
in Fig. \ref{fig6}(a). Unlike the earlier two boundaries where the edge
states still exist when $\delta =0$, the edge states disappear for armchair
boundary, which is similar to the situation of the fermionic lattice in
graphene \cite{Tan}. For $D\neq 0$, the degeneracy of those edge states is
slightly opened and can be directly connected with the non-zero Chern
number. These edge states always exist for $\delta \geqslant 0$, as shown in
Fig. \ref{fig6}(b), which implies the non-trivial topological properties.

\begin{figure}[tbp]
\begin{center}
\includegraphics[width=0.48\textwidth]{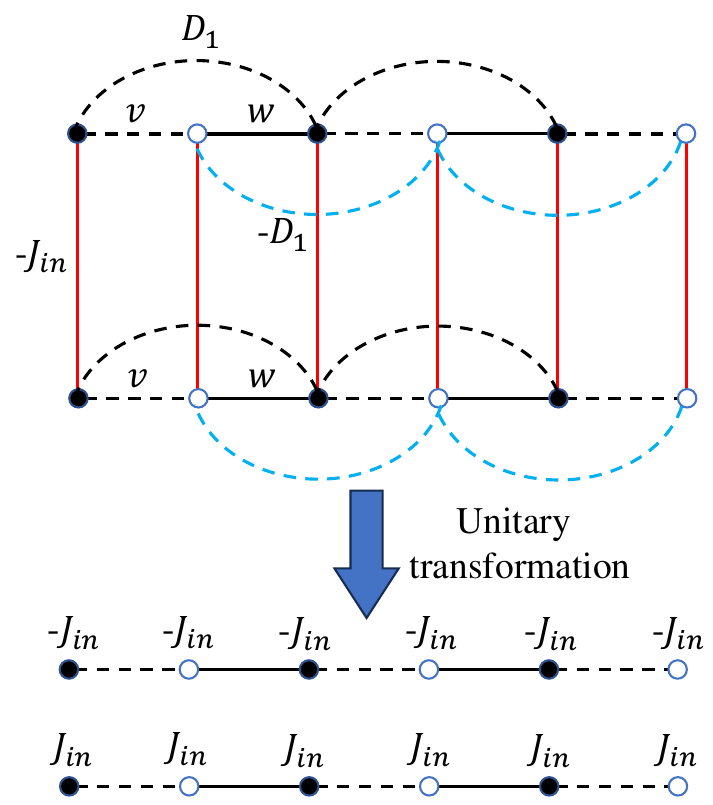}
\end{center}
\caption{(Color online) Top panel: The schematic illustration of quasi-1D
effective SSH ladder parameterized by $k_{y}$ for magnons in AA stacking
bilayer ferromagnets (FM interlayer interaction) on honeycomb lattices with
the bearded boundary shown in Fig. \protect\ref{fig2}(a). The on-site
potentials are omitted for simplicity with $w=-2SJ\cos (k_{y}/2)$, $v=-SJ$,
and $D_{1}=2SD\sin (k_{y}/2)$. Bottom panel: Two isolated SSH chains with
energy modulation $\pm J_{\mathrm{in}}$, which is equivalent to the SSH
ladder in the top panel. Here the next-nearest neighbor hopping $D_{1}$ is
omitted for simplicity.}
\label{fig7}
\end{figure}

\section{Topological classification and related edge states for magnons in
bilayer ferromagnets on honeycomb lattices}

\label{4}

\subsection{AA stacking}

For the bilayer situation, both the stacking form and the interlayer
interaction type between two layers can be different, which obviously
influence the physical properties of the magnon. Here we first consider AA
stacking form, i.e., two ferromagnets stack with each other without any
rotation or shifts. One can write the Hamiltonian as%
\begin{equation}
H=\sum_{l=1,2}H_{\mathrm{FM}}^{l}+H_{\mathrm{int}},
\end{equation}%
$l=1,2$ represents the layer index and the spin of $H_{\mathrm{FM}}^{1}$ and 
$H_{\mathrm{FM}}^{2}$ can be same or different. So there are two types of $%
H_{\mathrm{int}}$: the FM one and AFM one, as shown in Fig. \ref{fig1}(b).

\subsubsection{FM interlayer interaction}

Combining with the bulk BdG Hamiltonian $H_{\mathrm{FM}}^{l}$ with bearded
boundary as we showed in the last section as an example, the FM interlayer
interaction can be expressed as 
\begin{eqnarray}
H_{\mathrm{int}} &=&-J_{\mathrm{in}}\sum_{i}\boldsymbol{S}_{i}^{1}\cdot 
\boldsymbol{S}_{i}^{2} \\
&\approx &J_{\mathrm{in}}S\sum_{m=1}^{M}\sum_{n=1}^{N}\{\left[ \left(
a_{m,n}^{2}\right) ^{\dagger }a_{m,n}^{2}+\left( b_{m,n}^{2}\right)
^{\dagger }b_{m,n}^{2}\right]   \notag \\
&&+\left[ \left( a_{m,n}^{1}\right) ^{\dagger }a_{m,n}^{1}+\left(
b_{m,n}^{1}\right) ^{\dagger }b_{m,n}^{1}\right]   \notag \\
&&-\left[ \left( a_{m,n}^{1}\right) ^{\dagger }a_{m,n}^{2}+\left(
b_{m,n}^{1}\right) ^{\dagger }b_{m,n}^{2}\right] +H.c.\}  \notag
\end{eqnarray}%
in the basis of BdG Hamiltonian, with $J_{\mathrm{in}}>0$ is the strength of
the interlayer interaction. The straightforward calculation shows in
momentum space,%
\begin{eqnarray}
H_{\mathrm{int}}\left( \boldsymbol{k}\right)  &=&J_{\mathrm{in}}S[\left( a_{%
\boldsymbol{k}}^{2}\right) ^{\dagger }a_{\boldsymbol{k}}^{2}+\left( b_{%
\boldsymbol{k}}^{2}\right) ^{\dagger }b_{\boldsymbol{k}}^{2} \\
&&+\left( a_{\boldsymbol{k}}^{1}\right) ^{\dagger }a_{\boldsymbol{k}%
}^{1}+\left( b_{\boldsymbol{k}}^{1}\right) ^{\dagger }b_{\boldsymbol{k}}^{1}
\notag \\
&&-\left( a_{\boldsymbol{k}}^{1}\right) ^{\dagger }a_{\boldsymbol{k}%
}^{2}-\left( b_{\boldsymbol{k}}^{1}\right) ^{\dagger }b_{\boldsymbol{k}%
}^{2}+H.c.].  \notag
\end{eqnarray}%
It is easy to find the bulk magnonic BdG Hamiltonian of AA stacking
ferromagnets in momentum space has the form%
\begin{eqnarray*}
H\left( \boldsymbol{k}\right)  &=&\eta _{k}^{\dagger }H_{BdG}^{\mathrm{%
bea-bea}}\left( \boldsymbol{k}\right) \eta _{k},\eta _{k}=\left( \xi
_{k},\xi _{-k}^{\dagger }\right) ^{T}, \\
\xi _{k} &=&\left( a_{\boldsymbol{k}}^{1},b_{\boldsymbol{k}}^{1},a_{%
\boldsymbol{k}}^{2},b_{\boldsymbol{k}}^{2}\right) ,
\end{eqnarray*}%
with%
\begin{equation*}
H_{BdG}^{\mathrm{bea-bea}}\left( \boldsymbol{k}\right) =\left( 
\begin{array}{cc}
h\left( \boldsymbol{k}\right)  & 0 \\ 
0 & h^{\ast }\left( -\boldsymbol{k}\right) 
\end{array}%
\right) ,
\end{equation*}%
\begin{eqnarray}
h\left( \boldsymbol{k}\right)  &=&\frac{S}{2}[\left( \gamma +J_{\mathrm{in}%
}\right) \mathit{I}_{4}  \label{FM_bea} \\
&&-\left( 
\begin{array}{cc}
h_{1}\left( \boldsymbol{k}\right)  & \Delta  \\ 
\Delta  & h_{2}\left( \boldsymbol{k}\right) 
\end{array}%
\right) ],  \notag \\
h_{l}\left( \boldsymbol{k}\right)  &=&\left( 
\begin{array}{cc}
-\beta  & \alpha  \\ 
\alpha ^{\ast } & \beta 
\end{array}%
\right) ,\Delta =\left( 
\begin{array}{cc}
J_{\mathrm{in}} & 0 \\ 
0 & J_{\mathrm{in}}%
\end{array}%
\right) ,  \notag
\end{eqnarray}%
where $\alpha ,\beta $, and $\gamma $ are as same as in Eq. (\ref{m_bea}). A
more clarified form can be found by a simple gauge transformation as%
\begin{eqnarray}
h\left( \boldsymbol{k}\right)  &\rightarrow &G^{-1}h\left( \boldsymbol{k}%
\right) G \\
&=&\frac{S}{2}[\left( \gamma +J_{\mathrm{in}}\right) \mathit{I}_{4}  \notag
\\
&&-\left( 
\begin{array}{cc}
h_{1}\left( \boldsymbol{k}\right) +J_{\mathrm{in}} & 0 \\ 
0 & h_{2}\left( \boldsymbol{k}\right) -J_{\mathrm{in}}%
\end{array}%
\right) ],  \notag \\
G &=&\left( 
\begin{array}{cccc}
1 & 0 & 1 & 0 \\ 
0 & 1 & 0 & 1 \\ 
1 & 0 & -1 & 0 \\ 
0 & 1 & 0 & -1%
\end{array}%
\right) ,  \notag
\end{eqnarray}%
leading to the eigenenergy as%
\begin{equation*}
\varepsilon _{k}=\frac{S}{2}\left( \gamma +J_{\mathrm{in}}\pm J_{\mathrm{in}%
}\right) \pm \frac{S}{2}\left( \sqrt{\beta ^{2}+\left\vert \alpha
\right\vert ^{2}}\right) ,
\end{equation*}%
where the interlayer interaction coupling $J_{\mathrm{in}}$ plays as an
detuned energy factor, opening a gap $SJ_{\mathrm{in}}$ between magnonic
bands of one monolayer ferromagnet with the other. Here one should attention
that $\sqrt{\beta ^{2}+\left\vert \alpha \right\vert ^{2}}-J_{\mathrm{in}}$
can be zero due to the values of parameters $J$, $J_{\mathrm{in}}$ and $D$,
so the energies $\frac{S}{2}\left( \gamma +2J_{\mathrm{in}}-\sqrt{\beta
^{2}+\left\vert \alpha \right\vert ^{2}}\right) $ and $\frac{S}{2}\left(
\gamma +\sqrt{\beta ^{2}+\left\vert \alpha \right\vert ^{2}}\right) $ can be
degenerate in some specific parameters, meanwhile other two energies are
always gapped. Since the Hamiltonian has been diagonalized as two blocks,
the symmetry analysis can only focus on one of the blocks since $h_{1}\left( 
\boldsymbol{k}\right) =h_{2}\left( \boldsymbol{k}\right) $, which gives the
same classification as the monolayer case, as shown in Table \ref{Table I}.
The $%
%TCIMACRO{\U{2124} }%
%BeginExpansion
\mathbb{Z}
%EndExpansion
$ topological invariant is displayed by the related Chern number of four
eigenenergies%
\begin{eqnarray}
C_{1} &=&C_{2}=\left\{ 
\begin{array}{c}
1,D<0 \\ 
-1,D>0%
\end{array}%
\right. , \\
C_{3} &=&C_{4}=\left\{ 
\begin{array}{c}
-1,D<0 \\ 
1,D>0%
\end{array}%
\right. ,  \notag
\end{eqnarray}%
when all four energies are gapped. For the bulk Hamiltonian $H_{BdG}^{%
\mathrm{zig-zig}}\left( \boldsymbol{k}\right) $, it can be proved that $%
H_{BdG}^{\mathrm{zig-zig}}\left( \boldsymbol{k}\right) =$ $H_{BdG}^{\mathrm{%
bea-bea}}\left( \boldsymbol{k}\right) $ after a gauge transformation as we
did in the monolayer case.

Under this situation, the edge states of this bilayer ferromagnets with FM
interlayer coupling in AA stacking can be easily found from the edge states
of the monolayer case, where we display the SSH ladder chain parameterized
by $k_{y}$ for magnons in bilayer ferromagnets on honeycomb lattices with
the bearded boundary shown in Fig. \ref{fig2}(a) as an example to show this
point. It can be seen that in AA stacking, the SSH ladder chain
parameterized by $k_{y}$ can be turned into two isolated SSH chain with
different on-site potentials arising from the interlayer coupling. Each
chain provides a set of edge states as we shown in the monolayer case (Fig. %
\ref{fig3}), but with a energy modulation $J_{\mathrm{in}}\pm J_{\mathrm{in}%
} $, as shown in Fig. \ref{fig7}. The similar situation can be found for
other two types of boundaries.

\begin{figure}[tbp]
\begin{center}
\includegraphics[width=0.48\textwidth]{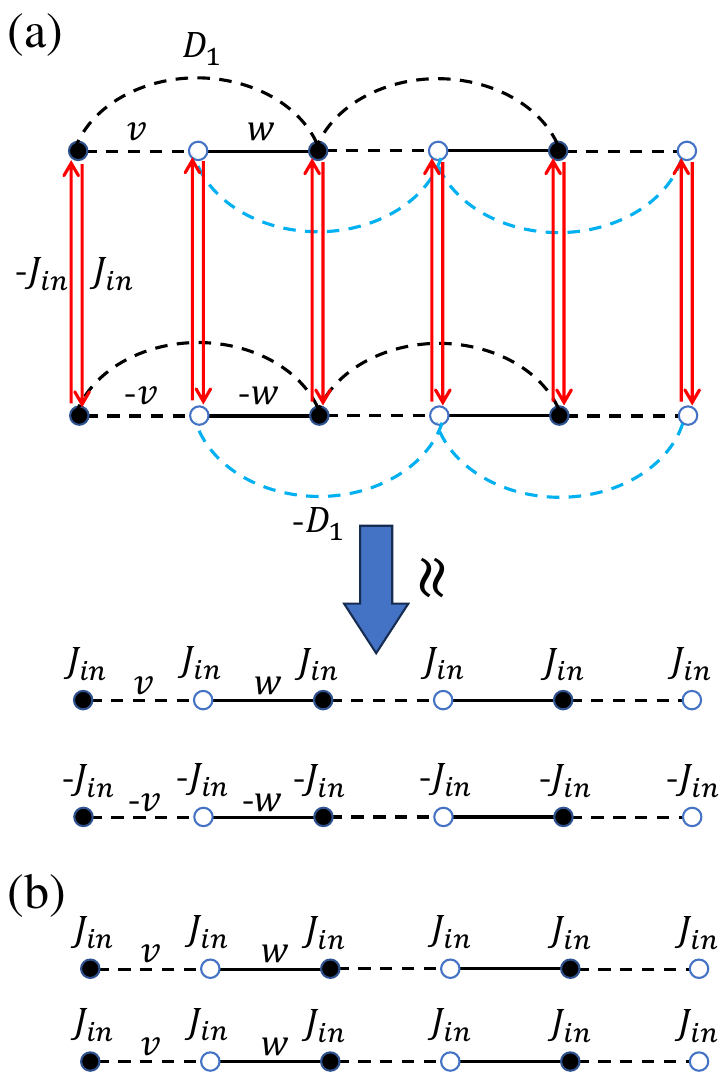}
\end{center}
\caption{(Color online) (a) Top panel: The schematic illustration of
quasi-1D effective SSH ladder $\protect\eta h(k_{y})$ for magnons in
AA-stacked bilayer ferromagnets (AFM interlayer interaction) on honeycomb
lattices with the bearded boundary shown in Fig. \protect\ref{fig2}(a). The
on-site potentials are omitted for simplicity with $w=-2SJ\cos (k_{y}/2)$, $%
v=-SJ$, and $D_{1}=2SD\sin (k_{y}/2)$. Bottom panel: Two isolated SSH chains
with energy modulation $\pm J_{\mathrm{in}}$, which is approximately
equivalent to the SSH ladder in the top panel. Here the next-nearest
neighbor hopping $D_{1}$ is omitted for simplicity. (b) The qualitative
schematic illustration of quasi-1D effective Hamiltonian $h(k_{y})$.}
\label{fig8}
\end{figure}

\subsubsection{AFM interlayer interaction}

The AFM one can be expressed as%
\begin{eqnarray}
H_{\mathrm{int}} &=&J_{\mathrm{in}}\sum_{i}\boldsymbol{S}_{i}^{1}\cdot 
\boldsymbol{S}_{i}^{2} \\
&\approx &J_{\mathrm{in}}S\sum_{m=1}^{M}\sum_{n=1}^{N}\{\left[ \left(
a_{m,n}^{2}\right) ^{\dagger }a_{m,n}^{2}+\left( b_{m,n}^{2}\right)
^{\dagger }b_{m,n}^{2}\right]   \notag \\
&&+\left[ \left( a_{m,n}^{1}\right) ^{\dagger }a_{m,n}^{1}+\left(
b_{m,n}^{1}\right) ^{\dagger }b_{m,n}^{1}\right]   \notag \\
&&-\left[ \left( a_{m,n}^{1}\right) ^{\dagger }\left( a_{m,n}^{2}\right)
^{\dagger }+\left( b_{m,n}^{1}\right) ^{\dagger }\left( b_{m,n}^{2}\right)
^{\dagger }\right] +H.c.\},  \notag
\end{eqnarray}%
with $J_{\mathrm{in}}>0$. Similar process as above shows that%
\begin{eqnarray}
H_{\mathrm{int}}\left( \boldsymbol{k}\right)  &=&J_{\mathrm{in}}S\{\left( a_{%
\boldsymbol{k}}^{2}\right) ^{\dagger }a_{\boldsymbol{k}}^{2}+\left( b_{%
\boldsymbol{k}}^{2}\right) ^{\dagger }b_{\boldsymbol{k}}^{2} \\
&&+\left( a_{\boldsymbol{k}}^{1}\right) ^{\dagger }a_{\boldsymbol{k}%
}^{1}+\left( b_{\boldsymbol{k}}^{1}\right) ^{\dagger }b_{\boldsymbol{k}}^{1}
\notag \\
&&-\left[ \left( a_{\boldsymbol{k}}^{1}\right) ^{\dagger }\left( a_{-%
\boldsymbol{k}}^{2}\right) ^{\dagger }+\left( b_{\boldsymbol{k}}^{1}\right)
^{\dagger }\left( b_{-\boldsymbol{k}}^{2}\right) ^{\dagger }\right] +H.c.\} 
\notag
\end{eqnarray}%
in momentum space and%
\begin{eqnarray}
H\left( \boldsymbol{k}\right)  &=&\eta _{k}^{\dagger }H_{BdG}\left( 
\boldsymbol{k}\right) \eta _{k},\eta _{k}=\left( \xi _{\boldsymbol{k}},\xi
_{-\boldsymbol{k}}^{\dagger }\right) ^{T}, \\
\xi _{\boldsymbol{k}} &=&\left[ a_{\boldsymbol{k}}^{1},b_{\boldsymbol{k}%
}^{1},\left( a_{-\boldsymbol{k}}^{2}\right) ^{\dagger },\left( b_{-%
\boldsymbol{k}}^{2}\right) ^{\dagger }\right] .  \notag
\end{eqnarray}%
Here we mainly focus on the bearded and zigzag boundaries for simplicity, it
gives%
\begin{equation*}
H_{BdG}^{\mathrm{bea-bea/zig-zig}}\left( \boldsymbol{k}\right) =\left( 
\begin{array}{cc}
h\left( \boldsymbol{k}\right)  & 0 \\ 
0 & h^{\ast }\left( -\boldsymbol{k}\right) 
\end{array}%
\right) ,
\end{equation*}%
which has the similar form with the FM case in Eq. (\ref{FM_bea}). However,
since the difference of spin between two layers and there is a coupling
between $\left( a_{\boldsymbol{k}}^{1}\right) ^{\dagger }$ and $\left( a_{-%
\boldsymbol{k}}^{2}\right) ^{\dagger }$ [$\left( b_{\boldsymbol{k}%
}^{1}\right) ^{\dagger }$ and $\left( b_{-\boldsymbol{k}}^{2}\right)
^{\dagger }$], the Hamiltonian%
\begin{equation}
H_{\sigma BdG}^{\mathrm{bea-bea/zig-zig}}\left( \boldsymbol{k}\right)
=\left( 
\begin{array}{cc}
\eta h\left( \boldsymbol{k}\right)  & 0 \\ 
0 & -\eta h^{\ast }\left( -\boldsymbol{k}\right) 
\end{array}%
\right) ,  \label{AFM_bea}
\end{equation}%
is no longer Hermitian with%
\begin{eqnarray}
&&\eta h\left( \boldsymbol{k}\right)  \\
&=&\frac{S}{2}\left( 
\begin{array}{cc}
\left( \gamma +J_{\mathrm{in}}\right) \mathit{I}_{2}-h_{1}\left( \boldsymbol{%
k}\right)  & -\Delta  \\ 
\Delta  & -\left( \gamma +J_{\mathrm{in}}\right) \mathit{I}_{2}+h_{2}\left( 
\boldsymbol{k}\right) 
\end{array}%
\right) ,  \notag
\end{eqnarray}%
where%
\begin{equation*}
h_{1}\left( \boldsymbol{k}\right) =\left( 
\begin{array}{cc}
\beta  & \alpha  \\ 
\alpha ^{\ast } & -\beta 
\end{array}%
\right) =h_{2}^{\ast }\left( -\boldsymbol{k}\right) .
\end{equation*}

Here, $\eta \equiv \sigma _{z}\otimes \mathit{I}_{2}$ can be seen as the pH
symmetric operator that we defined in the section \ref{2}. The eigenstates
of $\eta h\left( \boldsymbol{k}\right) $ are%
\begin{eqnarray*}
\left\vert \phi _{1}\right\rangle  &=&\frac{\left[ \alpha J_{\mathrm{in}%
}\left( A+\beta \right) ,-J_{\mathrm{in}}\lambda _{+}^{1},\alpha \lambda
_{+}^{2},-B_{+}\right] ^{T}}{\sqrt{\Omega _{+}}}\leftrightarrow -\varepsilon
_{+}, \\
\left\vert \phi _{2}\right\rangle  &=&\frac{\left[ \alpha J_{\mathrm{in}%
}\left( A+\beta \right) ,-J_{\mathrm{in}}\lambda _{-}^{1},\alpha \lambda
_{-}^{2},-B_{-}\right] ^{T}}{\sqrt{\Omega _{-}}}\leftrightarrow -\varepsilon
_{-}, \\
\left\vert \phi _{3}\right\rangle  &=&\frac{\left[ -B_{-},\alpha ^{\ast
}\lambda _{-}^{2},-J_{\mathrm{in}}\lambda _{-}^{1},\alpha ^{\ast }J_{\mathrm{%
in}}\left( A+\beta \right) \right] ^{T}}{\sqrt{\Omega _{-}}}\leftrightarrow
\varepsilon _{-}, \\
\left\vert \phi _{4}\right\rangle  &=&\frac{\left[ -B_{+},\alpha ^{\ast
}\lambda _{+}^{2},-J_{\mathrm{in}}\lambda _{+}^{1},\alpha ^{\ast }J_{\mathrm{%
in}}\left( A+\beta \right) \right] ^{T}}{\sqrt{\Omega _{+}}}\leftrightarrow
\varepsilon _{+},
\end{eqnarray*}%
with%
\begin{eqnarray*}
\varepsilon _{\pm } &=&[\left( \beta ^{2}+\left\vert \alpha \right\vert
^{2}\right) +A^{2}-J_{\mathrm{in}}^{2}\pm  \\
&&2\sqrt{\left( \beta ^{2}+\left\vert \alpha \right\vert ^{2}\right)
A^{2}-\beta ^{2}J_{\mathrm{in}}^{2}}]^{1/2},
\end{eqnarray*}%
\begin{eqnarray*}
\lambda _{\pm }^{1} &=&\beta \left( \beta -\varepsilon _{\pm }\right) \pm 
\sqrt{\left( \beta ^{2}+\left\vert \alpha \right\vert ^{2}\right)
A^{2}-\beta ^{2}J_{\mathrm{in}}^{2}}, \\
\lambda _{\pm }^{2} &=&A\left( A+\varepsilon _{\pm }\right) \pm \sqrt{\left(
\beta ^{2}+\left\vert \alpha \right\vert ^{2}\right) A^{2}-\beta ^{2}J_{%
\mathrm{in}}^{2}},
\end{eqnarray*}%
\begin{equation*}
A=\gamma +J_{\mathrm{in}},B_{\pm }=\frac{\left\vert \alpha \right\vert
^{2}\lambda _{\pm }^{2}-J_{in}^{2}\lambda _{\pm }^{1}}{\varepsilon _{\pm
}-\left( A-\beta \right) }.
\end{equation*}%
$\Omega _{\pm }$ are the normalization factors from the biorthogonal
relation. So the eigenenergies of $h\left( \boldsymbol{k}\right) $ can be
expressed as%
\begin{equation*}
\eta \left( 
\begin{array}{cccc}
\varepsilon _{+} & 0 & 0 & 0 \\ 
0 & \varepsilon _{-} & 0 & 0 \\ 
0 & 0 & -\varepsilon _{-} & 0 \\ 
0 & 0 & 0 & -\varepsilon _{+}%
\end{array}%
\right) =\left( 
\begin{array}{cccc}
\varepsilon _{+} & 0 & 0 & 0 \\ 
0 & \varepsilon _{-} & 0 & 0 \\ 
0 & 0 & \varepsilon _{-} & 0 \\ 
0 & 0 & 0 & \varepsilon _{+}%
\end{array}%
\right) 
\end{equation*}%
and the related eigenstates have the property as%
\begin{eqnarray*}
\left\langle \phi _{n}\right\vert \eta \left\vert \phi _{n}\right\rangle 
&=&1,n=3,4\allowbreak \allowbreak , \\
\left\langle \phi _{n}\right\vert \eta \left\vert \phi _{n}\right\rangle 
&=&-1,n=1,2\allowbreak \allowbreak , \\
\left\langle \phi _{n}\right\vert \eta \left\vert \phi _{m}\right\rangle 
&=&0,n\neq m,
\end{eqnarray*}%
where $\left\vert \phi _{n=1,2}\right\rangle $ and $\left\vert \phi
_{n=3,4}\right\rangle $ are not independent of each other and characterized
by the same topological invariant due to the symmetric restriction of the
free bosonic quadratic model \cite{Sat}. This means that although the Chern
number of $\left\vert \phi _{1}\right\rangle $ and $\left\vert \phi
_{2}\right\rangle $ should be different, i.e., $C_{1}+C_{2}=0$, $C_{1}=\pm
C_{4}$ and $C_{2}=\pm C_{3}$ should be satisfied, which has been proved by
our numerical calculation ($C_{1}=C_{4}$, $C_{2}=C_{3}$ in this case).
Therefore, for $D\neq 0$, one can easily found that $\eta h\left( 
\boldsymbol{k}\right) $ belongs to non-trivial topological class $\mathcal{A}
$ with $\eta $ (Table \ref{Table I}), and gives a $%
%TCIMACRO{\U{2124} }%
%BeginExpansion
\mathbb{Z}
%EndExpansion
$ topological invariant, shown by the related Chern number of $\eta h\left( 
\boldsymbol{k}\right) $ as 
\begin{equation}
C_{1}=\left\{ 
\begin{array}{c}
-1,D<0 \\ 
1,D>0%
\end{array}%
\right. .
\end{equation}

For the edge states of this bilayer ferromagnets with AFM interlayer
coupling in AA stacking under the open boundary condition, we still use the
honeycomb lattices with the bearded boundary shown in Fig. \ref{fig2}(a) as
an example. Using the effective SSH chain parameterized by $k_{y}$ for
magnons in monolayer case shown in Eq. (\ref{mbo}), the effective structure
parameterized by $k_{y}$ for magnons in this bilayer case can be expressed as%
\begin{equation}
H^{\mathrm{bea-bea}}\left( k_{y}\right) =\sum_{l=1,2}H_{l}^{\mathrm{bea}%
}\left( k_{y}\right) +H_{\mathrm{int}}\left( k_{y}\right) ,
\end{equation}%
with%
\begin{eqnarray}
&&H_{\mathrm{int}}\left( k_{y}\right) =J_{\mathrm{in}}S\sum_{n=1}^{N}\{\left[
\left( a_{k_{y},n}^{2}\right) ^{\dagger }a_{k_{y},n}^{2}+\left(
b_{k_{y},n}^{2}\right) ^{\dagger }b_{k_{y},n}^{2}\right]  \notag \\
&&+\left[ \left( a_{k_{y},n}^{1}\right) ^{\dagger }a_{k_{y},n}^{1}+\left(
b_{k_{y},n}^{1}\right) ^{\dagger }b_{k_{y},n}^{1}\right]  \notag \\
&&-\left[ \left( a_{k_{y},n}^{1}\right) ^{\dagger }\left(
a_{-k_{y},n}^{2}\right) ^{\dagger }+\left( b_{k_{y},n}^{1}\right) ^{\dagger
}\left( b_{-k_{y},n}^{2}\right) ^{\dagger }\right] +H.c.\}.  \notag
\end{eqnarray}%
Supposing $\xi _{k_{y}}^{p,l}=\left( p_{k_{y},1}^{l},\ldots
,p_{k_{y},N}^{l}\right) $, $p=a,b$, we have%
\begin{eqnarray*}
H^{\mathrm{bea-bea}}\left( k_{y}\right) &=&\eta _{k_{y}}^{\dagger
}H_{BdG}\left( k_{y}\right) \eta _{k_{y}},\eta _{k_{y}}=\left( \xi
_{k_{y}},\xi _{-k_{y}}^{\dagger }\right) ^{T} \\
\xi _{k_{y}} &=&\left[ \xi _{k_{y}}^{a,1},\xi _{k_{y}}^{b,1},\left( \xi
_{-k_{y}}^{a,2}\right) ^{\dagger },\left( \xi _{-k_{y}}^{b,2}\right)
^{\dagger }\right] ,
\end{eqnarray*}%
\begin{equation*}
H_{BdG}\left( k_{y}\right) =\left( 
\begin{array}{cc}
h\left( k_{y}\right) & 0 \\ 
0 & h^{\ast }\left( -k_{y}\right)%
\end{array}%
\right) ,
\end{equation*}%
giving the non-Hermitian Hamiltonian%
\begin{equation}
H_{\sigma BdG}\left( k_{y}\right) =\left( 
\begin{array}{cc}
\eta h\left( k_{y}\right) & 0 \\ 
0 & -\eta h^{\ast }\left( -k_{y}\right)%
\end{array}%
\right) ,
\end{equation}%
with $\eta =\sigma _{z}\otimes \mathit{I}_{2N}$. Interestingly, the
numerical calculation of the band structure proves that the effective
structure of $\eta h\left( k_{y}\right) $ in Fig. \ref{fig8} can be
approximately equivalent to two decoupled chains when $J_{\mathrm{in}}$ is
small compared with $J$ (always satisfied since it is an interlayer
interaction), instead of strictly equivalence in FM interlayer interaction
case. Therefore, the band structure of $h\left( k_{y}\right) $ is similar to
what we showed in Figs. \ref{fig3}(a) and (b) for different boundary
conditions, with every band is two-fold degenerate as explained in Fig. \ref%
{fig8}(b).

\begin{figure}[tbp]
\begin{center}
\includegraphics[width=0.48\textwidth]{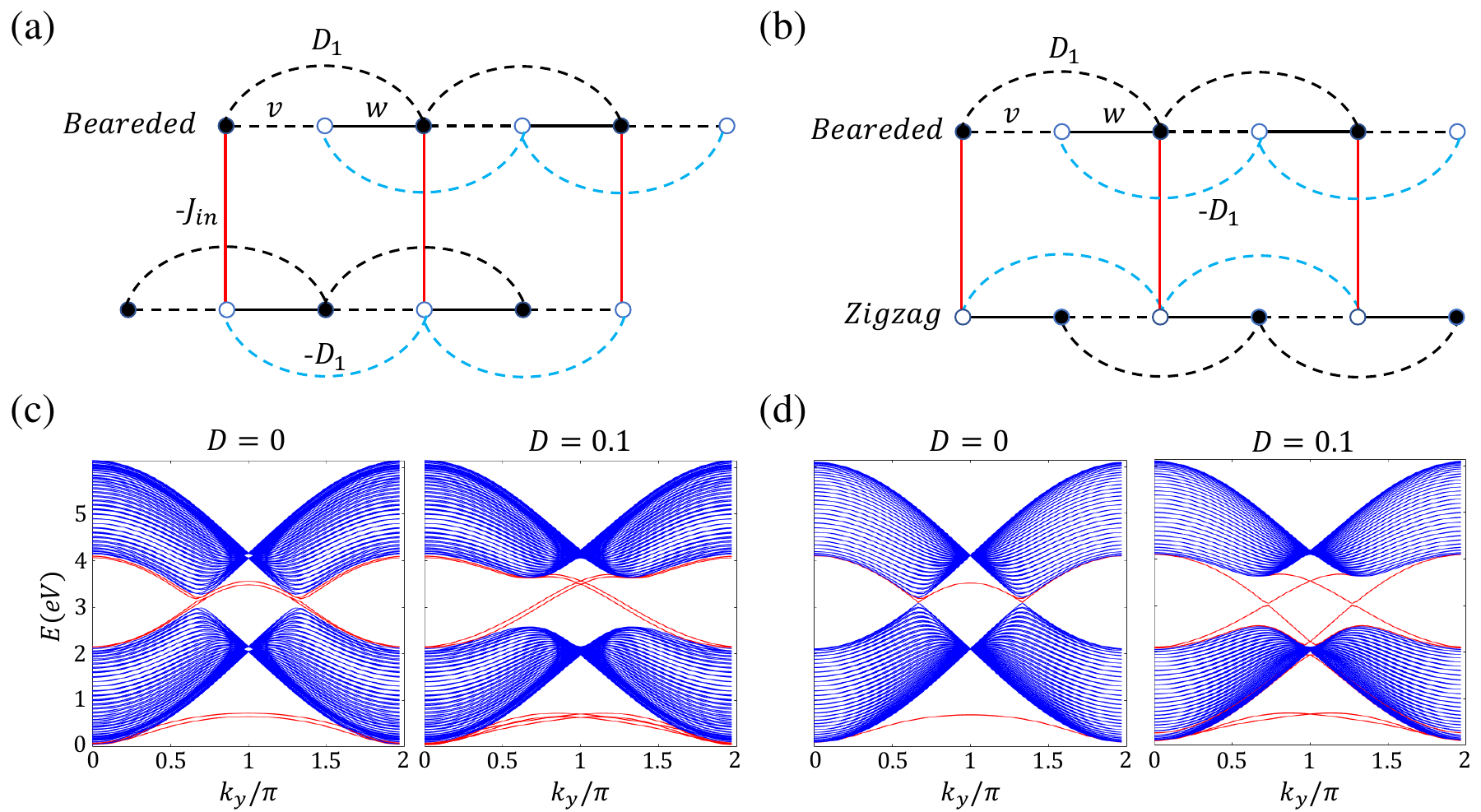}
\end{center}
\caption{(Color online) (a) The schematic illustration of quasi-1D effective
SSH ladder $h(k_{y})$ for magnons in AB stacking bilayer ferromagnets (FM
interlayer interaction) on honeycomb lattices with the bearded boundary
shown in Fig. \protect\ref{fig2}(a). The on-site potentials are omitted for
simplicity with $w=-2SJ\cos (k_{y}/2)$, $v=-SJ$, and $D_{1}=2SD\sin
(k_{y}/2) $. (b) The similar plot as (a) but with different boundaries in
top and bottom layers. (c) The band structure for effective chain in (a)
with different $D$ values (in the unit of $J$), where the edge states are
highlighted by the red solid lines. (d) The similar plot as (c) for
effective chain in (b). Here $S=1$, $J=1\,\mathrm{eV}$, $J_{\mathrm{in}%
}=0.1J $ and $\protect\gamma =3J$ ($\mathcal{K}=0$, $B_{z}=0$) is used for
all plots, respectively. }
\label{fig9}
\end{figure}

\subsection{AB stacking situation}

Here we consider AB stacking form, i.e., two ferromagnets stack with each
other by a relative shift. The Hamiltonian also has the form%
\begin{equation*}
H=\sum_{l=1,2}H_{\mathrm{FM}}^{l}+H_{\mathrm{int}}.
\end{equation*}

\subsubsection{FM interlayer interaction}

Still taking the bulk BdG Hamiltonian $H_{\mathrm{FM}}^{l}$ with the bearded
boundary condition that we showed in the last section as an example, the FM
interlayer interaction in this condition can be expressed as 
\begin{eqnarray}
H_{\mathrm{int}} &=&-J_{\mathrm{in}}\sum_{\left\langle i,j\right\rangle }%
\boldsymbol{S}_{i}^{1}\cdot \boldsymbol{S}_{j}^{2} \\
&\approx &J_{\mathrm{in}}S\sum_{m=1}^{M}\sum_{n=1}^{N}\{\left[ \left(
a_{m,n}^{2}\right) ^{\dagger }a_{m,n}^{2}+\left( b_{m,n}^{1}\right)
^{\dagger }b_{m,n}^{1}\right]  \notag \\
&&-\left( b_{m,n}^{1}\right) ^{\dagger }a_{m,n}^{2}+H.c.\}  \notag
\end{eqnarray}%
in the basis of BdG Hamiltonian with $J_{\mathrm{in}}>0$. The
straightforward calculation shows in momentum space,%
\begin{eqnarray}
H_{\mathrm{int}}\left( \boldsymbol{k}\right) &=&J_{\mathrm{in}}S[\left( a_{%
\boldsymbol{k}}^{2}\right) ^{\dagger }a_{\boldsymbol{k}}^{2}+\left( b_{%
\boldsymbol{k}}^{1}\right) ^{\dagger }b_{\boldsymbol{k}}^{1} \\
&&-\left( b_{\boldsymbol{k}}^{1}\right) ^{\dagger }a_{\boldsymbol{k}%
}^{2}+H.c.].  \notag
\end{eqnarray}%
Then the bulk magnonic BdG Hamiltonian of AB stacking ferromagnets in
momentum space has the form%
\begin{eqnarray*}
H\left( \boldsymbol{k}\right) &=&\eta _{k}^{\dagger }H_{BdG}^{\mathrm{bea-bea%
}}\left( \boldsymbol{k}\right) \eta _{k},\eta _{k}=\left( \xi _{k},\xi
_{-k}^{\dagger }\right) ^{T}, \\
\xi _{k} &=&\left( a_{\boldsymbol{k}}^{1},b_{\boldsymbol{k}}^{1},a_{%
\boldsymbol{k}}^{2},b_{\boldsymbol{k}}^{2}\right) ,
\end{eqnarray*}%
with%
\begin{equation*}
H_{BdG}^{\mathrm{bea-bea}}\left( \boldsymbol{k}\right) =\left( 
\begin{array}{cc}
h\left( \boldsymbol{k}\right) & 0 \\ 
0 & h^{\ast }\left( -\boldsymbol{k}\right)%
\end{array}%
\right) ,
\end{equation*}%
\begin{equation}
h\left( \boldsymbol{k}\right) =\frac{S}{2}\left[ \gamma \mathit{I}%
_{4}+\left( 
\begin{array}{cccc}
\beta & -\alpha & 0 & 0 \\ 
-\alpha ^{\ast } & J_{\mathrm{in}}-\beta & -J_{\mathrm{in}} & 0 \\ 
0 & -J_{\mathrm{in}} & J_{\mathrm{in}}+\beta & -\alpha \\ 
0 & 0 & -\alpha ^{\ast } & -\beta%
\end{array}%
\right) \right] .  \label{FM_AB_bea}
\end{equation}%
where $\alpha ,\beta $, and $\gamma $ are as same as in Eq. (\ref{m_bea}).
The eigenenergy can be expressed as

\begin{eqnarray}
\varepsilon _{1} &=&\frac{S}{2}\left( \gamma -\sqrt{\beta ^{2}+\left\vert
\alpha \right\vert ^{2}}\right) ,\varepsilon _{2}=\frac{S}{2}\left( \gamma
+J_{\mathrm{in}}-A\right) ,  \notag \\
\varepsilon _{3} &=&\frac{S}{2}\left( \gamma +\sqrt{\beta ^{2}+\left\vert
\alpha \right\vert ^{2}}\right) ,\varepsilon _{4}=\frac{S}{2}\left( \gamma
+J_{\mathrm{in}}+A\right) ,  \notag \\
A &=&\sqrt{J_{\mathrm{in}}^{2}+B^{2}},B=\sqrt{\beta ^{2}+\left\vert \alpha
\right\vert ^{2}},
\end{eqnarray}

with

\begin{eqnarray*}
\left\vert \phi _{1}\right\rangle &=&\frac{\left[ \alpha ^{2},\alpha \left(
\beta +B\right) ,\alpha \left( \beta +B\right) ,\left( \beta +B\right) ^{2} %
\right] ^{T}}{\sqrt{\Omega _{1}}}, \\
\left\vert \phi _{3}\right\rangle &=&\frac{\left[ \left( \beta +B\right)
^{2},-\alpha ^{\ast }\left( \beta +B\right) ,-\alpha ^{\ast }\left( \beta
+B\right) ,(\alpha ^{\ast })^{2}\right] ^{T}}{\sqrt{\Omega _{3}}},
\end{eqnarray*}%
\begin{eqnarray*}
\left\vert \phi _{2}\right\rangle &=&\frac{\left[ -\alpha ,J_{\mathrm{in}
}-A-\beta ,-\left( \beta +J_{\mathrm{in}}-A\right) ,\alpha ^{\ast }\right]
^{T}}{\sqrt{\Omega _{2}}}, \\
\left\vert \phi _{4}\right\rangle &=&\frac{\left[ -\alpha ,J_{\mathrm{in}
}+A-\beta ,-\left( \beta +J_{\mathrm{in}}+A\right) ,\alpha ^{\ast }\right]
^{T}}{\sqrt{\Omega _{4}}}.
\end{eqnarray*}

The symmetry analysis shows (see Table \ref{Table I}\ for details) $D\neq 0$
leads to a non-trivial topological class A and the $%
%TCIMACRO{\U{2124} }%
%BeginExpansion
\mathbb{Z}
%EndExpansion
$ topological invariant is shown as the related Chern number%
\begin{equation}
C_{2}=C_{4}=0,C_{1}=-C_{3}=\left\{ 
\begin{array}{c}
2,D<0 \\ 
-2,D>0%
\end{array}%
\right. .
\end{equation}

Moreover, one can have two samples with different boundaries to stack
together when considering the AB-stacking case on honeycomb lattices. If we
consider the bottom layer keeps the same boundary we used before (bearded)
and the top layer with a different boundary (zigzag), we have%
\begin{eqnarray*}
H\left( \boldsymbol{k}\right)  &=&\eta ^{\dagger }H_{BdG}^{\mathrm{bea-zig}%
}\left( \boldsymbol{k}\right) \eta ,\eta =\left( \xi _{k},\xi _{-k}^{\dagger
}\right) ^{T}, \\
\xi _{k} &=&\left( a_{\boldsymbol{k}}^{1},b_{\boldsymbol{k}}^{1},b_{%
\boldsymbol{k}}^{2},a_{\boldsymbol{k}}^{2}\right) 
\end{eqnarray*}%
\begin{equation*}
H_{BdG}^{\mathrm{bea-zig}}\left( \boldsymbol{k}\right) =\left( 
\begin{array}{cc}
h\left( \boldsymbol{k}\right)  & 0 \\ 
0 & h^{\ast }\left( -\boldsymbol{k}\right) 
\end{array}%
\right) ,
\end{equation*}%
\begin{equation}
h\left( \boldsymbol{k}\right) =\frac{S}{2}\left[ \gamma \mathit{I}%
_{4}+\left( 
\begin{array}{cccc}
J_{\mathrm{in}}+\beta  & -\alpha _{1} & -J_{\mathrm{in}} & 0 \\ 
-\alpha _{1}^{\ast } & -\beta  & 0 & 0 \\ 
-J_{\mathrm{in}} & 0 & J_{\mathrm{in}}-\beta  & -\alpha _{2}^{\ast } \\ 
0 & 0 & -\alpha _{2} & \beta 
\end{array}%
\right) \right] .
\end{equation}%
$\allowbreak \allowbreak \allowbreak \allowbreak $with $\alpha _{1}=\alpha
=J[1+e^{-ik_{x}}+e^{-i\left( k_{x}+k_{y}\right) }],\alpha
_{2}=J[1+e^{ik_{x}}+e^{-ik_{y}}]$.\ Through a gauge transformation%
\begin{eqnarray*}
\xi _{k}^{T} &\rightarrow &\left( 
\begin{array}{cccc}
0 & 0 & 0 & 1 \\ 
0 & 0 & e^{ik_{x}} & 0 \\ 
e^{ik_{x}} & 0 & 0 & 0 \\ 
0 & e^{ik_{x}} & 0 & 0%
\end{array}%
\right) \left( 
\begin{array}{c}
a_{\boldsymbol{k}}^{1} \\ 
b_{\boldsymbol{k}}^{1} \\ 
b_{\boldsymbol{k}}^{2} \\ 
a_{\boldsymbol{k}}^{2}%
\end{array}%
\right)  \\
&=&\left( 
\begin{array}{c}
a_{\boldsymbol{k}}^{2} \\ 
e^{ik_{x}}b_{\boldsymbol{k}}^{2} \\ 
e^{ik_{x}}a_{\boldsymbol{k}}^{1} \\ 
e^{ik_{x}}b_{\boldsymbol{k}}^{1}%
\end{array}%
\right) ,
\end{eqnarray*}%
one can easily find that%
\begin{equation}
h\left( \boldsymbol{k}\right) \rightarrow \frac{S}{2}\left[ \gamma \mathit{I}%
_{4}+\left( 
\begin{array}{cccc}
\beta  & -\alpha  & 0 & 0 \\ 
-\alpha ^{\ast } & J_{\mathrm{in}}-\beta  & -J_{\mathrm{in}} & 0 \\ 
0 & -J_{\mathrm{in}} & J_{\mathrm{in}}+\beta  & -\alpha  \\ 
0 & 0 & -\alpha ^{\ast } & -\beta 
\end{array}%
\right) \right] ,
\end{equation}%
which is the same bulk Hamiltonian in momentum space as before.

For the edge states of this bilayer ferromagnets with FM interlayer coupling
in AB stacking under the open boundary condition, we use the honeycomb
lattices with the bearded and zigzag boundaries shown in Figs. \ref{fig2}(a)
and (b) as examples. According to the effective SSH chain parameterized by $%
k_{y}$ for magnons in monolayer case shown in Eq. (\ref{mbo}), the effective
structure parameterized by $k_{y}$ for magnons in this bilayer case can be
expressed as in Fig. \ref{fig9}. The related band structures indicate that
although the bulk Hamiltonians for conditions in Figs. \ref{fig9}(a) and (b)
are the same, i.e., they have the same bulk topological properties, the
distribution of edge states in $k_{y}$ space is totally different, as shown
in Figs. \ref{fig9}(c) and (d). This is related with different topological
properties of the quasi-1D effective structure parameterized by $k_{y}$ for
magnons, which can be measured by the winding number similar to the electric
system in graphene \cite{Tan}.

\subsubsection{AFM interlayer interaction}

The AFM one can be expressed as%
\begin{eqnarray}
H_{\mathrm{int}} &=&J_{\mathrm{in}}\sum_{i}\boldsymbol{S}_{i}^{1}\cdot 
\boldsymbol{S}_{i}^{2} \\
&\approx &J_{\mathrm{in}}S\sum_{m=1}^{M}\sum_{n=1}^{N}\{\left[ \left(
a_{m,n}^{2}\right) ^{\dagger }a_{m,n}^{2}+\left( b_{m,n}^{1}\right)
^{\dagger }b_{m,n}^{1}\right]   \notag \\
&&-\left( b_{m,n}^{1}\right) ^{\dagger }\left( a_{m,n}^{2}\right) ^{\dagger
}+H.c.\},  \notag
\end{eqnarray}%
with%
\begin{equation*}
H_{\mathrm{int}}\left( \boldsymbol{k}\right) =J_{\mathrm{in}}S[\left( a_{%
\boldsymbol{k}}^{2}\right) ^{\dagger }a_{\boldsymbol{k}}^{2}+\left( b_{%
\boldsymbol{k}}^{1}\right) ^{\dagger }b_{\boldsymbol{k}}^{1}-\left( b_{%
\boldsymbol{k}}^{1}\right) ^{\dagger }\left( a_{-\boldsymbol{k}}^{2}\right)
^{\dagger }+H.c.].
\end{equation*}%
in momentum space. Then%
\begin{eqnarray}
H\left( \boldsymbol{k}\right)  &=&\eta ^{\dagger }H_{BdG}^{\mathrm{bea-bea}%
}\left( \boldsymbol{k}\right) \eta ,\eta =\left( \xi _{\boldsymbol{k}},\xi
_{-\boldsymbol{k}}^{\dagger }\right) ^{T}, \\
\xi _{\boldsymbol{k}} &=&\left[ a_{\boldsymbol{k}}^{1},b_{\boldsymbol{k}%
}^{1},\left( a_{-\boldsymbol{k}}^{2}\right) ^{\dagger },\left( b_{-%
\boldsymbol{k}}^{2}\right) ^{\dagger }\right] .  \notag
\end{eqnarray}%
Still focus on the bearded and zigzag boundaries for simplicity, it gives%
\begin{equation*}
H_{BdG}^{\mathrm{bea-bea}}\left( \boldsymbol{k}\right) =\left( 
\begin{array}{cc}
h\left( \boldsymbol{k}\right)  & 0 \\ 
0 & h^{\ast }\left( -\boldsymbol{k}\right) 
\end{array}%
\right) ,
\end{equation*}%
\begin{equation}
h\left( \boldsymbol{k}\right) =\frac{S}{2}\left[ \gamma \mathit{I}%
_{4}-\left( 
\begin{array}{cccc}
\beta  & \alpha  & 0 & 0 \\ 
\alpha ^{\ast } & -J_{\mathrm{in}}-\beta  & J_{\mathrm{in}} & 0 \\ 
0 & J_{\mathrm{in}} & -J_{\mathrm{in}}-\beta  & \alpha  \\ 
0 & 0 & \alpha ^{\ast } & \beta 
\end{array}%
\right) \right] .  \label{AFM_AB_bea}
\end{equation}%
The same step before results in non-Hermitian Hamiltonian%
\begin{equation}
H_{\sigma BdG}^{\mathrm{bea-bea}}\left( \boldsymbol{k}\right) =\left( 
\begin{array}{cc}
\eta h\left( \boldsymbol{k}\right)  & 0 \\ 
0 & -\eta h^{\ast }\left( -\boldsymbol{k}\right) 
\end{array}%
\right) ,
\end{equation}%
and $\eta \equiv \sigma _{z}\otimes \mathit{I}_{2}$ still means the pH
symmetric operator. $H_{\sigma BdG}^{\mathrm{bea-bea}}\left( \boldsymbol{k}%
\right) \simeq $ $H_{\sigma BdG}^{\mathrm{bea-zig}}\left( \boldsymbol{k}%
\right) $ can be from a gauge transformation as we proved before. The
topological classification in Table \ref{Table I} shows that it has the same
classification as the AA stacking case. The eigenstates for $\eta h\left( 
\boldsymbol{k}\right) $ are%
\begin{eqnarray*}
\left\vert \phi _{1}\right\rangle  &=&\frac{\left( \alpha ,\lambda
_{+}^{2},\lambda _{+}^{1},\alpha ^{\ast }\frac{\lambda _{+}^{1}}{%
-\varepsilon _{+}+\gamma -\beta }\right) ^{T}}{\sqrt{\Omega _{+}}}%
\leftrightarrow -\varepsilon _{+}, \\
\left\vert \phi _{2}\right\rangle  &=&\frac{\left( \alpha ,\lambda
_{-}^{2},\lambda _{-}^{1},\alpha ^{\ast }\frac{\lambda _{-}^{1}}{%
-\varepsilon _{-}+\gamma -\beta }\right) ^{T}}{\sqrt{\Omega _{-}}}%
\leftrightarrow -\varepsilon _{-}, \\
\left\vert \phi _{3}\right\rangle  &=&\frac{\left( \alpha \frac{\lambda
_{-}^{1}}{-\varepsilon _{-}+\gamma -\beta },\lambda _{-}^{1},\lambda
_{-}^{2},\alpha ^{\ast }\right) ^{T}}{\sqrt{\Omega _{-}}}\leftrightarrow
\varepsilon _{-}, \\
\left\vert \phi _{4}\right\rangle  &=&\frac{\left( \alpha \frac{\lambda
_{+}^{1}}{-\varepsilon _{-}+\gamma -\beta },\lambda _{+}^{1},\lambda
_{+}^{2},\alpha ^{\ast }\right) ^{T}}{\sqrt{\Omega _{+}}}\leftrightarrow
\varepsilon _{+},
\end{eqnarray*}%
\begin{eqnarray*}
A &=&-J_{in}\left( \beta +\gamma \right) -2\beta \gamma ,B=\sqrt{%
A^{2}+4\left\vert \alpha \right\vert ^{2}\gamma \left( \gamma +J_{in}\right) 
}, \\
\varepsilon _{\pm } &=&\frac{1}{\sqrt{4\gamma \left( \gamma +J_{in}\right) }}%
\{[\sqrt{A^{2}+4\left\vert \alpha \right\vert ^{2}\gamma \left( \gamma
+J_{in}\right) } \\
&&\pm 2\gamma \left( \gamma +J_{in}\right) ]^{2}-J_{in}^{2}\left( \gamma
-\beta \right) ^{2}\}^{1/2}, \\
\lambda _{\pm }^{1} &=&\frac{2\gamma \left( \gamma +\varepsilon _{\pm
}+J_{in}\right) +J_{in}\varepsilon _{\pm }\pm B}{J_{in}},\lambda _{\pm
}^{2}=-\beta +\gamma +\varepsilon _{\pm },
\end{eqnarray*}%
giving the eigenenergies of $h\left( k\right) $ can be expressed as%
\begin{equation*}
\eta \left( 
\begin{array}{cccc}
\varepsilon _{+} & 0 & 0 & 0 \\ 
0 & \varepsilon _{-} & 0 & 0 \\ 
0 & 0 & -\varepsilon _{-} & 0 \\ 
0 & 0 & 0 & -\varepsilon _{+}%
\end{array}%
\right) =\left( 
\begin{array}{cccc}
\varepsilon _{+} & 0 & 0 & 0 \\ 
0 & \varepsilon _{-} & 0 & 0 \\ 
0 & 0 & \varepsilon _{-} & 0 \\ 
0 & 0 & 0 & \varepsilon _{+}%
\end{array}%
\right) .
\end{equation*}%
The related eigenstates have the property as%
\begin{eqnarray*}
\left\langle \phi _{n}\right\vert \left( \sigma _{z}\otimes \mathit{I}%
_{2}\right) \left\vert \phi _{n}\right\rangle  &=&1,n=3,4\allowbreak
\allowbreak , \\
\left\langle \phi _{n}\right\vert \left( \sigma _{z}\otimes \mathit{I}%
_{2}\right) \left\vert \phi _{n}\right\rangle  &=&-1,n=1,2\allowbreak
\allowbreak , \\
\left\langle \phi _{n}\right\vert \left( \sigma _{z}\otimes \mathit{I}%
_{2}\right) \left\vert \phi _{m}\right\rangle  &=&0,n\neq m,
\end{eqnarray*}%
which is similar to the AA stacking case with AFM interlayer interaction.
This leads to the related Chern number of $\eta h\left( \boldsymbol{k}%
\right) $ as 
\begin{equation}
C_{1}=\left\{ 
\begin{array}{c}
-1,D<0 \\ 
1,D>0%
\end{array}%
\right. ,
\end{equation}%
with $C_{1}+C_{2}=0$, $C_{1}=-C_{4}$, $C_{2}=-C_{3}$.

\begin{figure}[tbp]
\begin{center}
\includegraphics[width=0.48\textwidth]{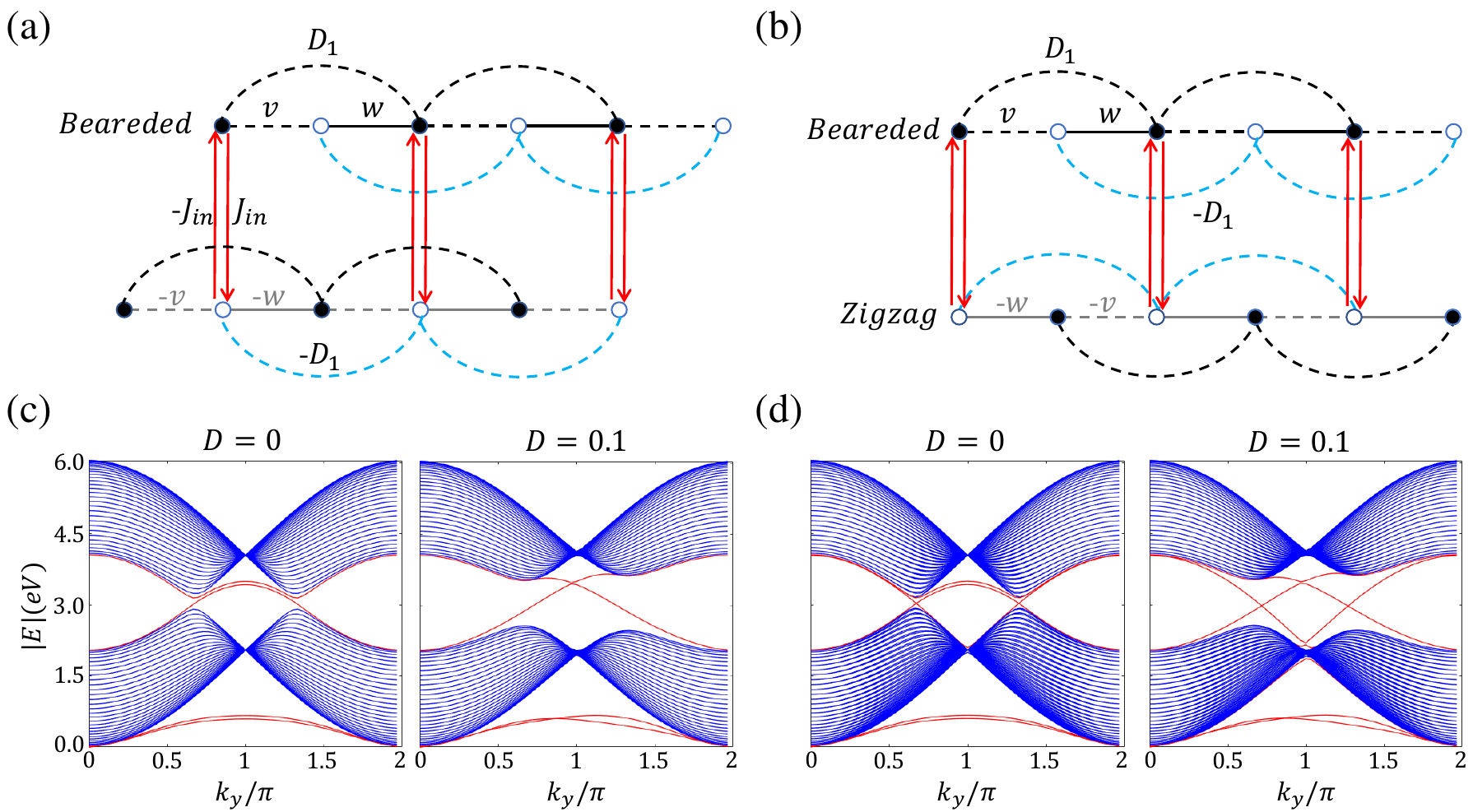}
\end{center}
\caption{(Color online) (a) The schematic illustration of quasi-1D effective
SSH ladder $\protect\eta h(k_{y})$ for magnons in AB stacking bilayer
ferromagnets (AFM interlayer interaction) on honeycomb lattices with the
bearded boundary shown in Fig. \protect\ref{fig2}(a). The on-site potentials
are omitted for simplicity with $w=-2SJ\cos (k_{y}/2)$, $v=-SJ$, and $%
D_{1}=2SD\sin (k_{y}/2)$. (b) The similar plot as (a) but with different
boundaries in top and bottom layers. (c) The absolute value of the band
structure for effective chain in (a) with different $D$ values (in the unit
of $J$), where the edge states are highlighted by the red solid lines. (d)
The similar plot as (c) for effective chain in (b). Here $S=1$, $J=1\,%
\mathrm{eV}$, $J_{\mathrm{in}}=0.1J$ and $\protect\gamma =3J$ ($\mathcal{K}%
=0 $, $B_{z}=0$) is used for all plots, respectively. }
\label{fig10}
\end{figure}

For the edge states of this bilayer ferromagnets with AFM interlayer
coupling in AB stacking under the open boundary condition, as we derived for
AA stacking case, it can be described by the band structure of an effective
Hamiltonian $\eta h\left( k_{y}\right) $. The results are shown in Fig. \ref%
{fig10}, which the similar band structure to the AB stacking case with FM
interlayer interaction. Here one should attention that we use the absolute
value of energy ($\left\vert E\right\vert $) to show this the band structure
of $\eta h\left( k_{y}\right) $ for clarity, which is equivalent to the band
structure $\left\vert E\right\vert $ of $h\left( k_{y}\right) $. Meanwhile,
in details, the band degeneracy of both the bulk and edge bands are opened
in FM case for the bearded boundary condition, but strictly keeping for AFM
case with the same condition, as shown in Figs. \ref{fig9}(c) and \ref{fig10}%
(c). If the boundary conditions for top and bottom layers are different, the
situation is reversed for FM and AFM cases, as shown in Figs. \ref{fig9}(d)
and \ref{fig10}(d). Also, when $D\neq 0$, the shape of edge bands can be
influenced by the AFM interlayer interaction, which is no longer symmetric
in $k_{y}$ space as shown in Figs. \ref{fig9}(c) and (d). This can be
connected to the non-Hermitian properties of Hamiltonian $\eta h\left(
k_{y}\right) $ that imaginary energies appear (which is pretty small
compared to the real part, but non-zero, so we show $\left\vert E\right\vert 
$ in Fig. \ref{fig10} for simplicity).

\section{Summary}

In summary, we study the topological classification and related edge states
of magnons in ferromagnets on honeycomb that can be described by a class of
single-particle bosonic BdG models. Both single layer and bilayer situations
are considered. For the monolayer case, since there are no double
annihilation or creation operators in the related BdG models, the
conventional bulk-edge correspondence for Hermitian systems remains
effective. The non-zero Chern number is directly connected with non-zero
DMI, but its relation with the specific band can be influenced by the
boundary condition. Non-trivial edge states can emerge even without the DMI
under certain open boundary conditions. For the bilayer case, the stacking
form (AA or AB) and the type of interlayer coupling, i.e., ferromagnetic
(FM) or antiferromagnetic (AFM), can significantly alter the Chern number
for magnon bands. Additionally, the different boundary conditions for
different layers play a crucial role in the distribution of edge states in
momentum space.

%The present attempt to identify the fundamental roles played by adiabatic and non-adiabatic interband transitions in Hall conduction also merits in that it offer a non-geometric approach to understand the non-Abelian characters of the gauge structures from multiple bands

\textit{Acknowledgments.}---C. Li would like to thank D. W. Zhai and B. Fu
for useful discussions. The work is mainly supported by the Research Grants
Council of Hong Kong (HKU17306819 and AoE/P-701/20), and the University of
Hong Kong (Seed Funding for Strategic Interdisciplinary Research).

%\section*{References}

\end{document}